\documentclass[useAMS,usenatbib]{mn2e}
\usepackage[pdftex]{graphicx}
\usepackage{setspace}
\usepackage{color,soul}

\title[Detection of water in HAT-P-1b]{HST hot Jupiter Transmission Spectral Survey: detection of water in HAT-P-1b from WFC3 near-IR spatial scan observations}

\author[H. R. Wakeford et al.]{H. R. Wakeford$^1$\thanks{hannah@astro.ex.ac.uk},
D.~K.~Sing$^1$, 
D.~Deming$^2$, 
N. P.~Gibson$^3$,
J. J.~Fortney$^4$,   \newauthor 
A.~S.~Burrows$^5$,  
G. Ballester$^6$,
N. Nikolov$^1$, 
S. Aigrain$^{7}$, 
G. Henry$^{8}$,  \newauthor 
H. Knutson$^9$, 
A. Lecavelier des Etangs$^{10}$, 
 F. Pont$^1$,  
A.P. Showman$^6$, \newauthor  
A. Vidal-Madjar$^{10}$, 
K. Zahnle$^{11}$ 
 \\
$^1$ Astrophysics Group, School of Physics, University of Exeter, Stocker Road, Exeter EX4 4QL, UK\\
$^2$ Department of Astronomy, University of Maryland, College Park, MD 20742 USA \\
$^3$ ESO Karl-Schwarzschild-Strasse 2, D-85748 Garching bei M\"unchen\\
$^4$ Department of Astronomy and Astrophysics, University of California, Santa Cruz, CA 95064 USA \\
$^5$ Department of Astrophysical Sciences, Princeton University, Princeton, NJ 08544-1001 USA \\
$^6$ Lunar and Planetary Lab, University of Arizona, Tucson, AZ 85721 USA \\
$^{7}$ Department of Physics, University of Oxford, Denys Wilkinson Building, Keble Road, Oxford, OX1 3RH\\
$^{8}$ Tennessee State University, Nashville, TN 37203-3401 \\
$^9$ Division of Geological and Planetary Sciences, California Institute of Technology, Pasadena, CA 91125 USA \\
$^{10}$ Institut d'Astrophysique de Paris, CNRS, 98 bis Boulevard Arago, F-75014 Paris, France \\
$^{11}$ NASA Ames Research Center, Moffett Field, California 94035 }

\date{in original form 2013 April 4th}

\pagerange{\pageref{firstpage}--\pageref{lastpage}}\pubyear{2012}

\begin{document}

\label{firstpage}

\maketitle
%
%
\begin{abstract}
	\indent{} We present Hubble Space Telescope near-infrared transmission spectroscopy of the transiting hot-Jupiter HAT-P-1b. We observed one transit with Wide Field Camera 3 using the G141 low-resolution grism to cover the wavelength range 1.087-1.678\,$\mu$m. These time series observations were taken with the newly available spatial scan mode that increases the duty cycle by nearly a factor of two, thus improving the resulting photometric precision of the data. We measure a planet-to-star radius ratio of R$_{p}$/R$_{*}$=0.11709$\pm$0.00038 in the white light curve with the centre of transit occurring at 2456114.345$\pm$0.000133 (JD). We achieve S/N levels per exposure of 1840 (0.061\,\%) at a resolution of  $\Delta$$\lambda$=19.2\,nm (R$\sim$70) in the 1.1173 - 1.6549\,$\mu$m spectral region, providing the precision necessary to probe the transmission spectrum of the planet at close to the resolution limit of the instrument. We compute the transmission spectrum using both single target and differential photometry with similar results. The resultant transmission spectrum shows a significant absorption above the 5-$\sigma$ level matching the 1.4\,$\mu$m water absorption band. In solar composition models, the water absorption is sensitive to the $\sim$ 1\,mbar pressure levels at the terminator. The detected absorption agrees with that predicted by an 1000\,K isothermal model, as well as with that predicted by a planetary-averaged temperature model. 
\end{abstract}


%
%
\section{Introduction}
	\indent{}The understanding of exoplanetary atmospheres has advanced considerably in the last decade, thanks in part to the spectroscopic observations of transiting exoplanets. During a transit, when a planet passes between Earth and its host star, a small fraction of the starlight is blocked by the planet; this can then be seen as a characteristic dip in the transit light curve. Transiting planets offer a unique opportunity to study their atmospheres through a method called transmission spectroscopy. As the starlight passes through their upper atmospheres characteristic spectral signatures are superimposed on the light as it is absorbed or scattered. The absorption and optical depth of the atmosphere is dependant on wavelength, as is the altitude at which the planet is opaque to starlight. Features observed in the transmission spectrum place strong constraints on the possible species in the atmosphere (e.g. \citealt{seager2000}, \citealt{charbonneau2002}). 
	 
	\indent{} A range of atomic and molecular species have been identified in exoplanetary atmospheres through transmission spectroscopy, with a majority having been identified in the upper and lower atmospheres of HD 189733b and HD 209458b, which remain the most studied exoplanets to date. Ground- and space-based observations ranging from the ultraviolet (UV) to the infrared (IR) have been able to probe both the lower and extended upper atmosphere of these two exoplanets (for example: \citealt{vidal2003}, \citeyear{vidal2004}; \citealt{narita2005}; \citealt{pont2007}; \citealt{tinetti2007}; \citealt{snellen2008}; \citealt{redfield2008}; \citealt{swain2008}; \citealt{grillmair2008}; \citealt{desert2009}; \citealt{linsky2010}; \citealt{sing2011b}; \citealt{lecavelier2012}; \citealt{gibson2012b}; \citealt{deming2013}; \citealt{benjaffel2013}; \citealt{waldmann2013}). \\
	\indent{} H$_{2}$O is a key molecule for constraining hot-Jupiter atmospheres. It is predicted that the C/O ratio plays a pivotal role in the relative abundances of H$_{2}$O and the other spectroscopically important  CH$_{4}$, CO, CO$_{2}$, C$_{2}$H$_{4}$, and HCN molecules in the atmospheres of close-in giant planets (e.g. \citealt{seager2000}; \citealt{madhusudhan2012}). \citet{moses2012} have analysed transit and eclipse observations of a number of transiting hot Jupiters, finding that some extrasolar giant planets could have unexpectedly low abundance of H$_{2}$O due to high C/O ratios.  Atmospheres with solar elemental abundances in thermochemical equilibrium are expected to have abundant water vapour, and disequilibrium processes like photochemistry are not able to deplete water sufficiently in the infrared photosphere of these planets to explain the observations (see \citealt{moses2012} and references there in). Extinction from clouds and hazes could also significantly mask absorption signatures of water, however, this would also mask other molecular species making emission spectra appear more like a blackbody (\citealt{fortney2005}; \citealt{pont2012}). \\
	\indent{} In this paper we present the transmission spectrum of HAT-P-1b based on one transit observation between 1.1\,$\mu$m and 1.7\,$\mu$m using Hubble Space Telescope (HST) Wide Field Camera 3 (WFC3) in spatial scan mode.  HST/WFC3 IR observations at 1.1-1.7\,$\mu$m probe primarily the H$_{2}$O absorption band at 1.4\,$\mu$m. These observations are among the first results from a large survey with HST probing the transmission spectra, from the optical to near-IR, of eight hot-Jupiter exoplanets (GO program 12473, P.I. D. Sing). HAT-P-1b is a low-density hot Jupiter orbiting a single member of a visual stellar binary (\citealt{bakos2007}). HAT-P-1b orbits its host star with a period of 4.5 days at a distance of 0.055\,AU. It has a radius similar to that of HD\,209458b with a somewhat lower mean density with a mass of 0.54\,M$_{J}$.
			Spitzer IRAC secondary eclipse measurements show that the atmosphere is best fit with a modest temperature inversion with a maximum dayside temperature of 1550\,K, assuming zero albedo, a uniform temperature over the dayside hemisphere, and no transport to the nightside (\citealt{todorov2010}). Ks-band secondary-eclipse observations have also been conducted by the GROUnd-based Secondary Eclipse (GROUSE) project with an estimated brightness temperature of 2136$\pm$150\,K and for an eclipse depth of 0.109$\pm$0.025$\%$ although there are still visible systematics that remain in the fit (\citealt{demooij2011}). \\	
	 In \S2 we outline the observations and the use of spatial scan mode, in \S3 we present the analysis of the extracted light curves, in \S4 we compare the result with atmospheric models, and \S5 we state our conclusions. 

\begin{figure}
\begin{center}
	\includegraphics[width=8.5cm]{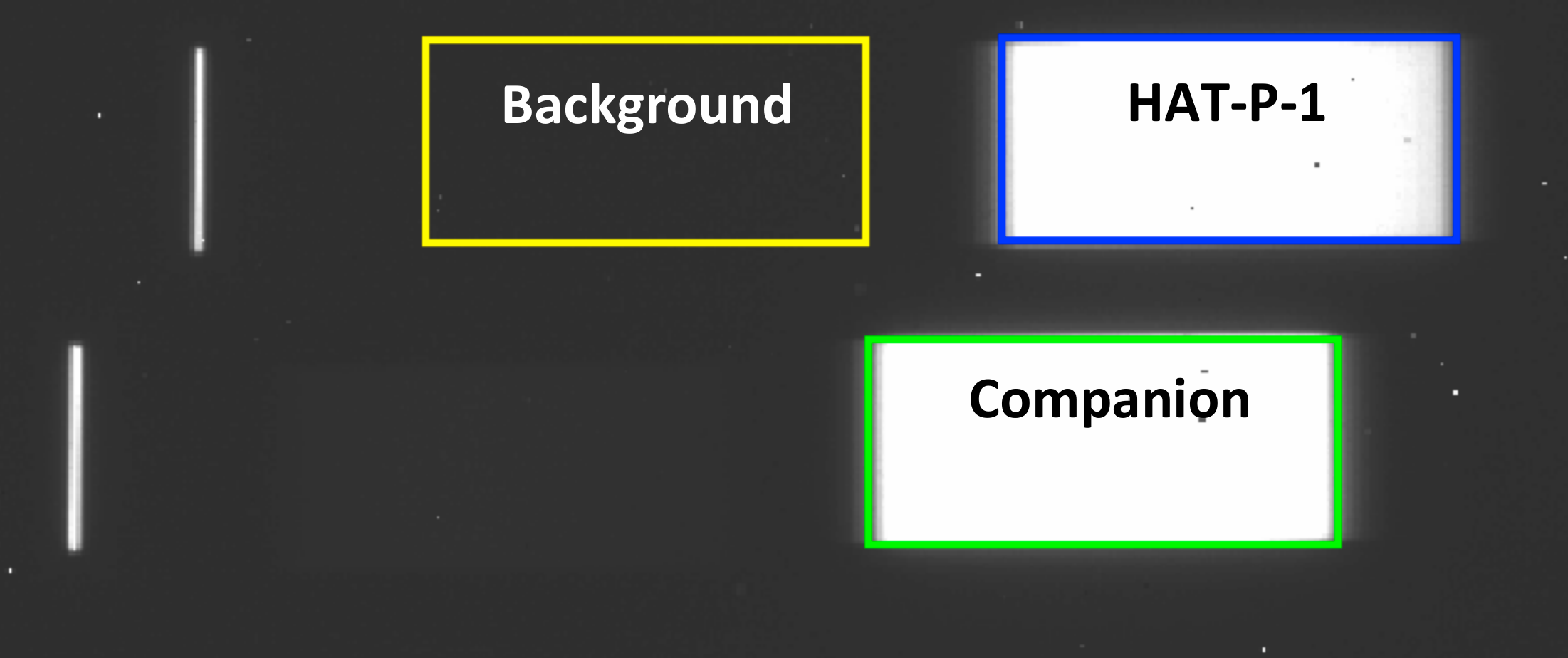}
	\end{center}
\caption{Cut-out of WFC3 G141 grism exposure with the spatial-scan spectra of the HAT-P-1 extraction window outlined in blue (top) and the G0 stellar companion outlined in green (bottom). To the left of HAT-P-1's spectra is the background subtraction region used in SPEXTRACT (outlined in a yellow box). }
\label{fig:imads9}
\end{figure}

%
\section{Observations}
	\indent{} Observations of HAT-P-1 were conducted in the NIR with HST/WFC3. WFC3's IR channel consists of a 1024x1024 pixel Teledyne HgCdTe detector that can be paired with any of 15 filters or two low-resolution grisms (\citealt{dressel2010}). Each exposure is compiled from multiple non-destructive reads (NSAMP) at either the full array or a subarray. 
	Although the standard WFC3 configuration is not particularly efficient for high S/N time series data, as buffer dumps are long and the PSF covers very few pixels (low S/N per exposure), the instrumental systematics are noticeably lower than for NICMOS as WFC3 does not suffer from strong intra-pixel sensitivities. WFC3 also has a factor of two improvement  on sensitivity over NICMOS with a much higher throughput and lower read-noise (e.g. WFC3 Instrument Handbook). 	
				
	\indent{}The observations started on July 5th, 2012 at 15:17 using the IR G141 grism in spatial scan mode over five HST orbits. We gathered exposures using 512 x 512 pixel subarrays with an NSAMP=4 readout sequence and exposure times of 46.69 seconds. 
	
\begin{figure}
\begin{center}
	\includegraphics[width=8.6cm]{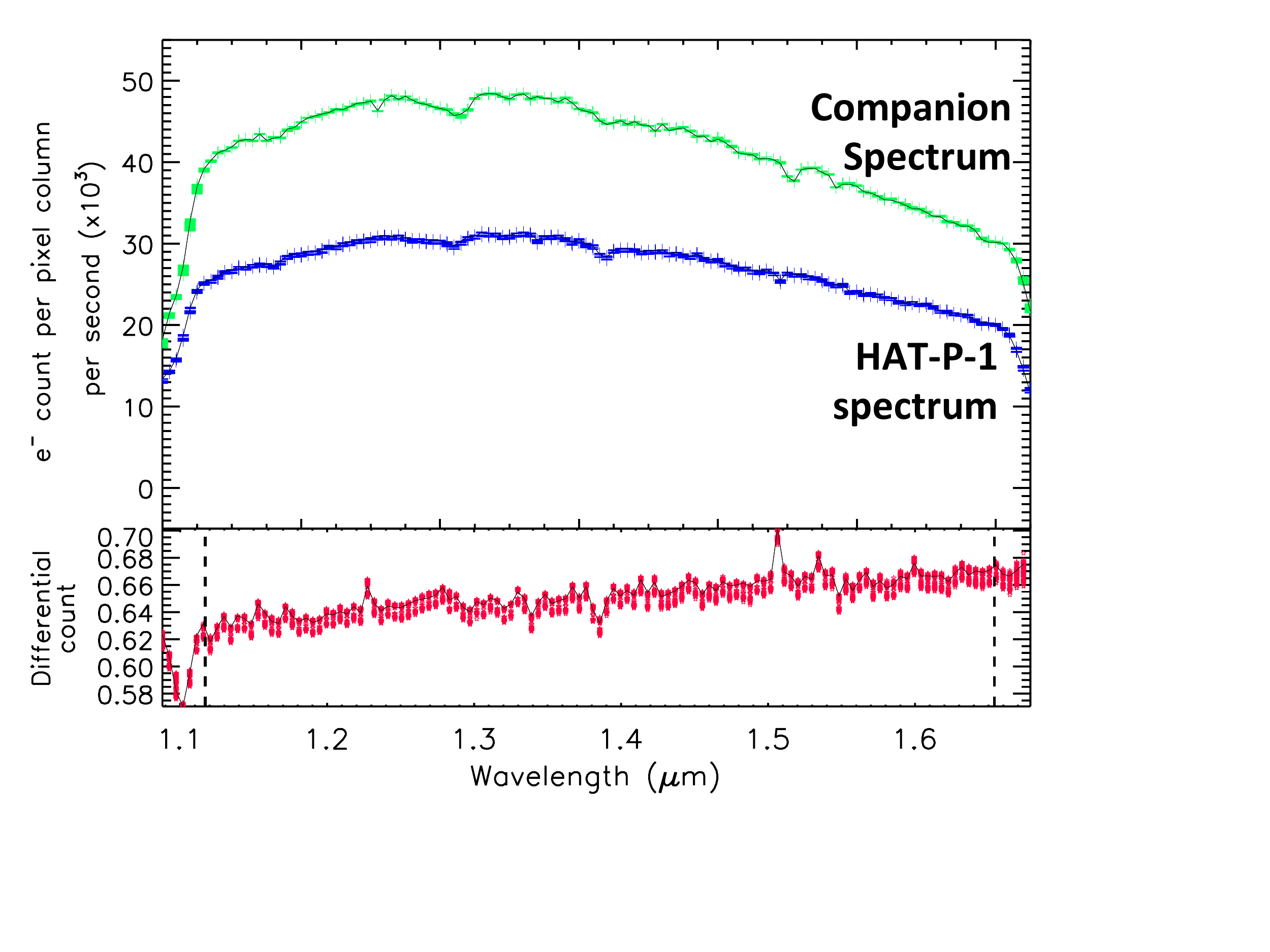}
\end{center}
\caption{Top: Spectra extracted from HST/WFC3 ``ima" images for HAT-P-1 (blue-lower) and its G0 binary companion (green-upper). Bottom: The resultant spectrum from differential photometric analysis; the vertical dashed lines define the wavelength range used in the spectroscopic analysis.}
\label{fig:spec}
\end{figure}

	\indent{} HAT-P-1 is the dimmer member of a double G0/G0 star system, ADS 16402, separated by 11.2$^{\prime\prime}$ (\citealt{bakos2011}).  Both stars are clearly resolved in the 68$^{\prime\prime}$ x 68$^{\prime\prime}$ field of view of HST/WFC3's spatial-scan spectra and are easily extracted separately in the analysis (see Fig. \ref{fig:imads9} and \ref{fig:spec}). This provides the opportunity to perform differential photometry on HAT-P-1 using the companion's signal which can reduce observational systematics in the data (see Fig. \ref{fig:white} and \ref{fig:whitend}).
	
\subsection{Spatial Scanning}
	\indent{} We present some of the first results from WFC3 using the spatial-scan mode to observe exoplanetary transits. The WFC3 spatial scanning involves nodding the telescope during an exposure to spread the light along the cross-dispersion axis, resulting in a higher number of photons by a factor of ten per exposure while considerably reducing overheads. This also increases the time of saturation of the brightest pixels, and allows for longer exposure times (\citealt{mccullough2011}). Our observations were conducted with a scan rate of 1.07 pixels per second, where 1 pixel = 0.13 arcseconds and thus spanning $\sim$50 pixels over each 46.69 second exposure. The duty cycle of the observations improved from 26\% in non-spatial-scan mode to 40\%.\\
\indent{}  The raw light curves of some WFC3 non-spatial-scan observations (e.g. \citealt{berta2012}) have been dominated by a systematic increase in intensity during each group of exposures obtained between buffer dumps referred to as the `hook' effect. It has been found that the `hook' is, on average, zero when the count rate is less than about 30,000 electrons per pixel (\citealt{deming2013}). We observe a maximum raw count rate of 25,000 electrons per pixel in our target star and a rate of $\sim$30,000 electrons per pixel for the companion star with no evidence for a significant `hook' effect in the reduced data of either star (see Fig. \ref{fig:white}).

%
\section{Analysis}
	\indent{} We used the ``{\it ima}" outputs from WFC3's \emph{Calwf3} pipeline. For each exposure, \emph{Calwf3} conducts the following processes: bad pixel flagging, reference pixel subtraction, zero-read subtraction, dark current subtraction, non-linearity correction, flat-field correction, as well as  gain and photometric calibration.  The resultant images are in units of electrons per second. \\
	\indent{} Subsequent data analysis is conducted with the first orbit removed (26 exposures), as it suffers from thermal breathing systematic effects that require time to settle, all previous transit studies have used a similar strategy (\citealt{brown2001b}; \citealt{charbonneau2002}; \citealt{sing2011b}). This leaves 86 exposures over the remaining 4 orbits with a total of 30 in transit exposures. The mid-time of each exposure was converted into BJD$_{TBD}$ for use in the transit light curves. \\
		\indent{} We used a box around each spectral image shown in Fig. \ref{fig:imads9}. The spectra were extracted using custom IDL procedures, similar to IRAF's APALL procedure, using an aperture of $\pm$23 pixels from the central row, determined by minimising the standard deviation across the aperture.
		 This 47-pixel aperture is slightly shorter than the total height of the spectrum to utilise pixels having similar exposure levels to the maximum possible degree.
		 The aperture is traced around a computed centring profile, which was found to be consistent in the y-axis within an error of 0.01 pixels. Background subtraction was applied using the region to the left of the HAT-P-1 spectrum (shown in Fig. \ref{fig:imads9}), because the region above and below each spectrum contains significant count levels which added noise to the resultant spectrum. 
		
\subsection{Wavelength Calibration}
	\indent{} For wavelength calibration, direct images were taken in the F139M narrow band filter at the beginning of the observations for a reference of the absolute position (X$_{ref}$, Y$_{ref}$) of the target star.
	We assumed that all pixels in the same column have the same effective wavelength, as the spatial scan varied in the $X_{ref}$ by less than one pixel, giving a spectral range of 1.087 - 1.678\,$\mu$m. 
		This wavelength range was later restricted to 1.1173 - 1.6549\,$\mu$m for the spectroscopic light curve fits as the strongly sloped edges covered by the grism response exhibit greater wavelength jitter where the intensities increase towards the edge of the bandpass (see Fig. \ref{fig:spec}).
		
	To calculate the wavelength corresponding to each pixel along the x-direction, we applied a linear fit to the wavelength solution. The wavelength solution is a function of the $X_{ref}$ and $Y_{ref}$ position given by, 
	$$ \lambda(x) = a0 + a1 \times X_{ref} $$
	and
	\begin{equation} \label{wavelength}
		 \lambda(pixel) = \lambda(x)\ +\ (Y_{ref\_dispersion} \times  X_{Pixel}) 
	\end{equation}
	where, $X_{ref}$ is taken from the filter image, $a0$ and $a1$ are taken from Table 5 in \citet{kuntschner2009}, and $Y_{ref\_dispersion}$ is found in Figure 6 of \citet{kuntschner2009} using the $Y_{ref}$ position from the filter image. \\
	\indent{} The G141 grism images contain both the 0th order, and the 1st order spectra for both stars. Each 1st order spectrum spans 128 pixels with a dispersion of 4.65 nm/pixel and the separation between the two stellar spectra was 23 pixels in the y-axis and 33 pixels in the x-axis (see Fig. \ref{fig:imads9}).\\	
\indent{} Using the zeroth order spectrum, we characterised the shift in Y$_{ref}$ over the course of the observations to monitor any shift in wavelength of the spectral trace. We observed a $\pm$0.2-pixel column shift in the wavelength direction over the whole observing period. This corresponded to 0.00186\,$\mu$m or a $\sim$10\% wavelength shift for each spectral bin over the span of the observations. We therefore adjusted the wavelength solution to use the average wavelength of the visit for each spectral bin. 
The observations, however, were relatively insensitive to sub-pixel wavelength shifts while the water spectral band spans a much larger wavelength range. \\
\indent{} Larger wavelength shifts were observed by \citet{deming2013} over the course of their observations of planetary transits which also revealed evidence of undersampling of the grism resolution by the pixel grid changing gradually and smoothly as a function of wavelength shift. To determine if our data contained similar undersampling, we compared a number of the spectral lines from the start and end of the observations (separated by over 3 hours) at a number of positions along the scanned spectra. Unlike the results found by \citet{deming2013} we see no flattening of the strong Paschen-beta stellar line at 1.28\,$\mu$m due to an undersampling effect. To help reduce the effects of any unidentified undersampling, we moderately binned our spectra effectively smoothing out any undersampling inherent in our data. 

\begin{figure} 
\begin{center}
	\includegraphics[width=8.5cm]{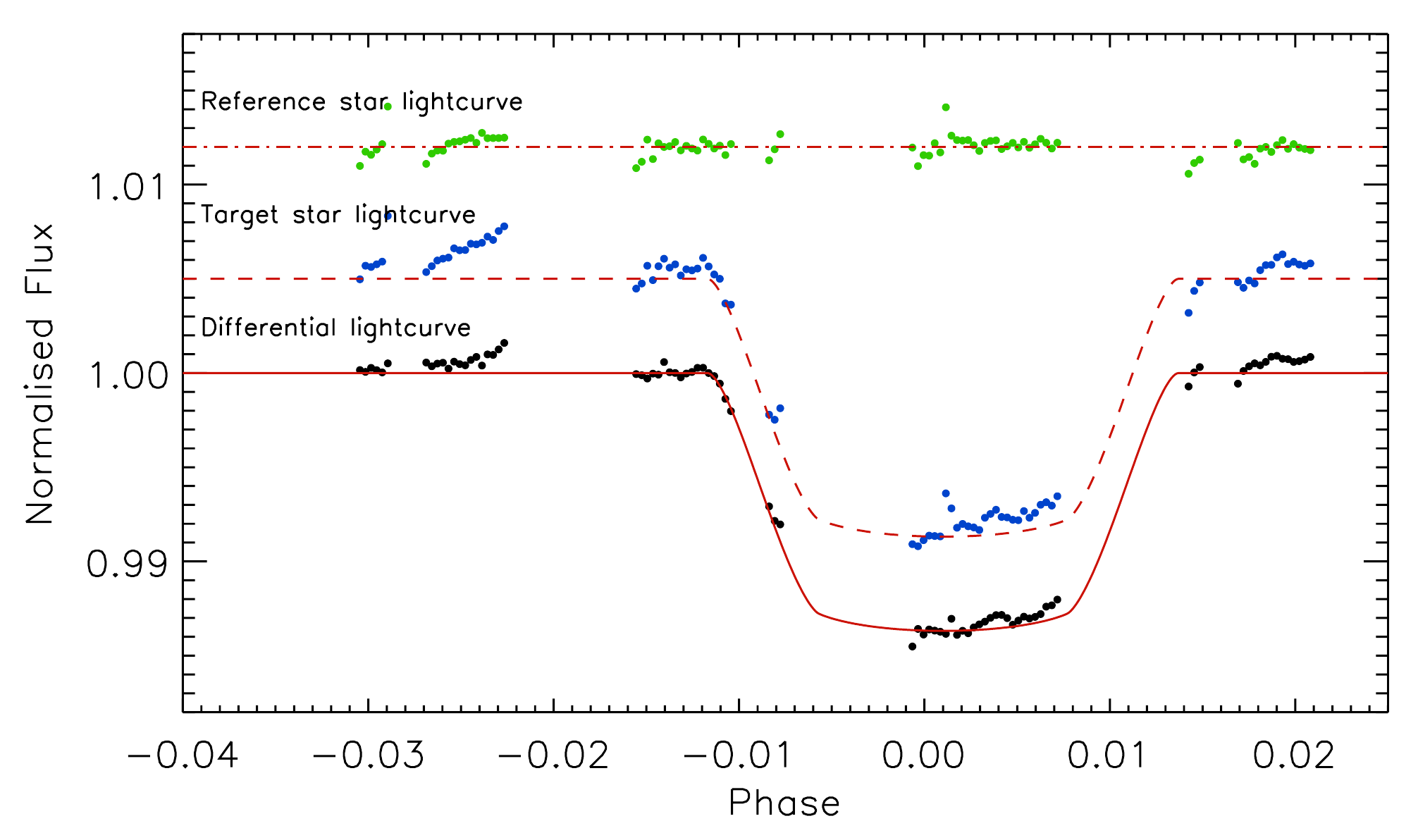}
\end{center}
\caption{The raw white light curve for the reference and target star as well as the raw differential light curve produced by dividing the target star light curve by the reference star light curve. Over-plotted in red are the Mandel and Agol (2002) limb-darkened transit models. The different light curves have been artificially shifted for clarity.}
\label{fig:whitend}
\end{figure}

\subsection{Limb Darkening} 
	\indent{} To accurately model the transit light curves, stellar limb darkening has to be carefully considered. The light curves were fit using the \citet{mandelagol2002} limb-darkened analytic transit model. We calculated limb-darkening coefficients from a 3D time dependent hydrodynamical model (\citealt{hayek2012}) over the wavelength range 1.1 - 1.7\,$\mu$m with the coefficients calculated separately for each spectral band. We also computed the limb-darkening coefficients using Kurucz stellar models for a star at T$_{eff}$=6000\,K, log g = 4.5, and [Fe/H] = +0.1 (\citealt{torres2008}). The coefficients were calculated following \citet{sing2010} using a non-linear limb-darkening law given by,
	\begin{equation}
	 \frac{I(\mu)}{I(1)} = 1 -  \sum_{n=1}^{4} c_{n}(1 - \mu^{\frac{n}{2}}) $$
	\end{equation}
	 where {\it I}(1) is the intensity at the centre of the stellar disk and $\mu = cos(\theta)$ is the angle between the line of sight and the emergent intensity. \\
	 \indent{} The 3D model shows overall weaker limb-darkening compared to the 1D model (\citealt{hayek2012}). The 3D model takes into account convective motions in the stellar atmosphere resulting in a shallower vertical temperature profile. As the strength of limb-darkening is closely related to the vertical atmospheric temperature gradient near the optical surface, the limb darkening slightly weakens for the shallower temperature profile. We find that this leads to an overall common shift in the derived planet-to-star radius ratio, with the shape of the transmission spectrum unaffected. We adopt the 3D model as it provides an overall better fit between our STIS (Space Telescope Imaging Spectrograph) and WFC3 data (\citealt{nikolov2013}). 
\begin{figure} 
\begin{center}
	\includegraphics[width=8.5cm]{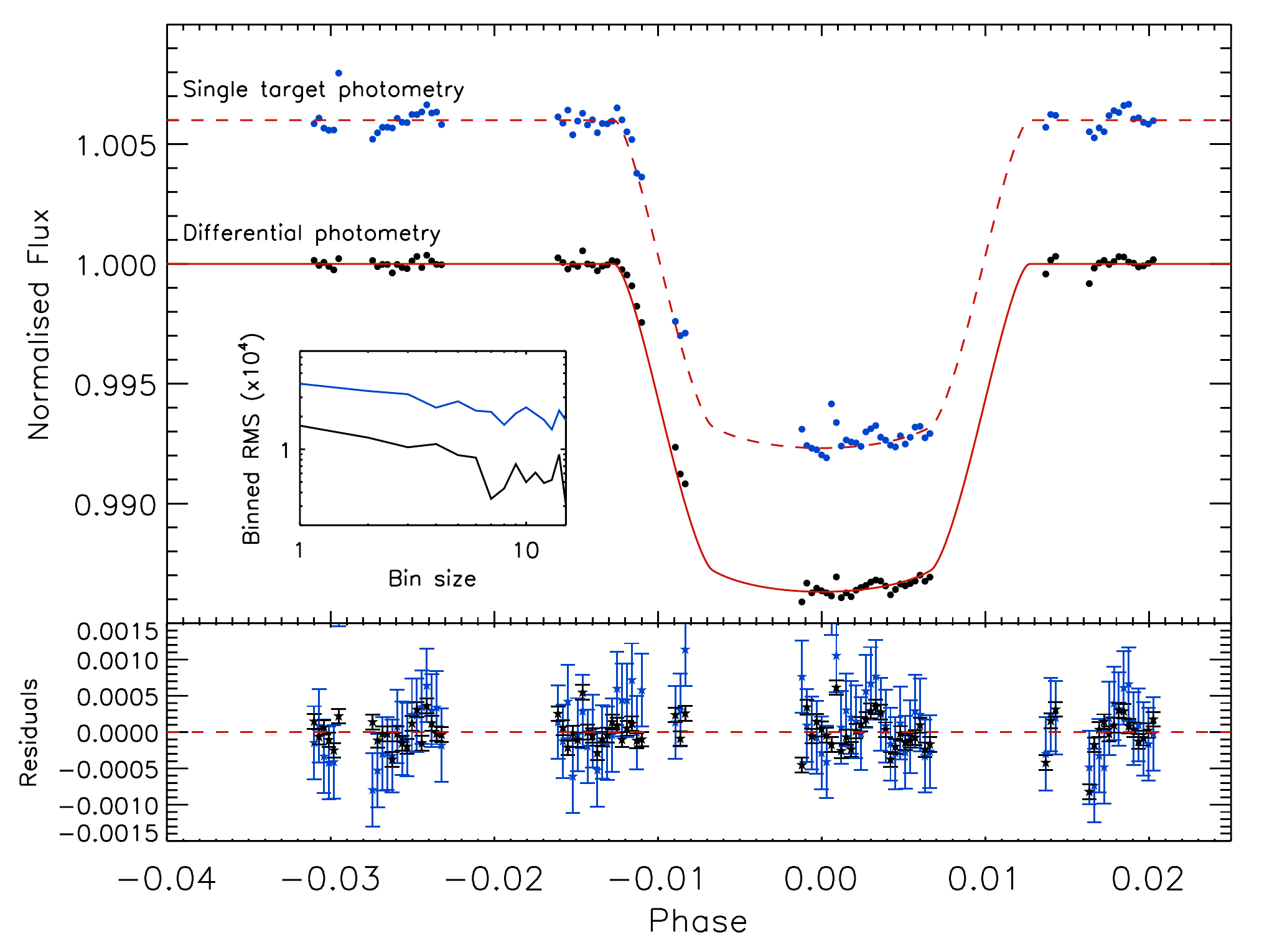}
\end{center}
\caption{Upper: Breathing-corrected light curves for both single target photometry (top curve) and differential photometry (bottom curve). The subplot shows the red noise for both single target (blue) and differential photometry (black) showing that time correlated noise is decreased when differential photometry is performed.  Lower: Corresponding residuals for both fits showing the decrease in errors and deviation from the mean when applying differential photometry to the data.}
\label{fig:white}
\end{figure}

\begin{figure*} 
\begin{center}
	\includegraphics[width=18cm, height=13cm]{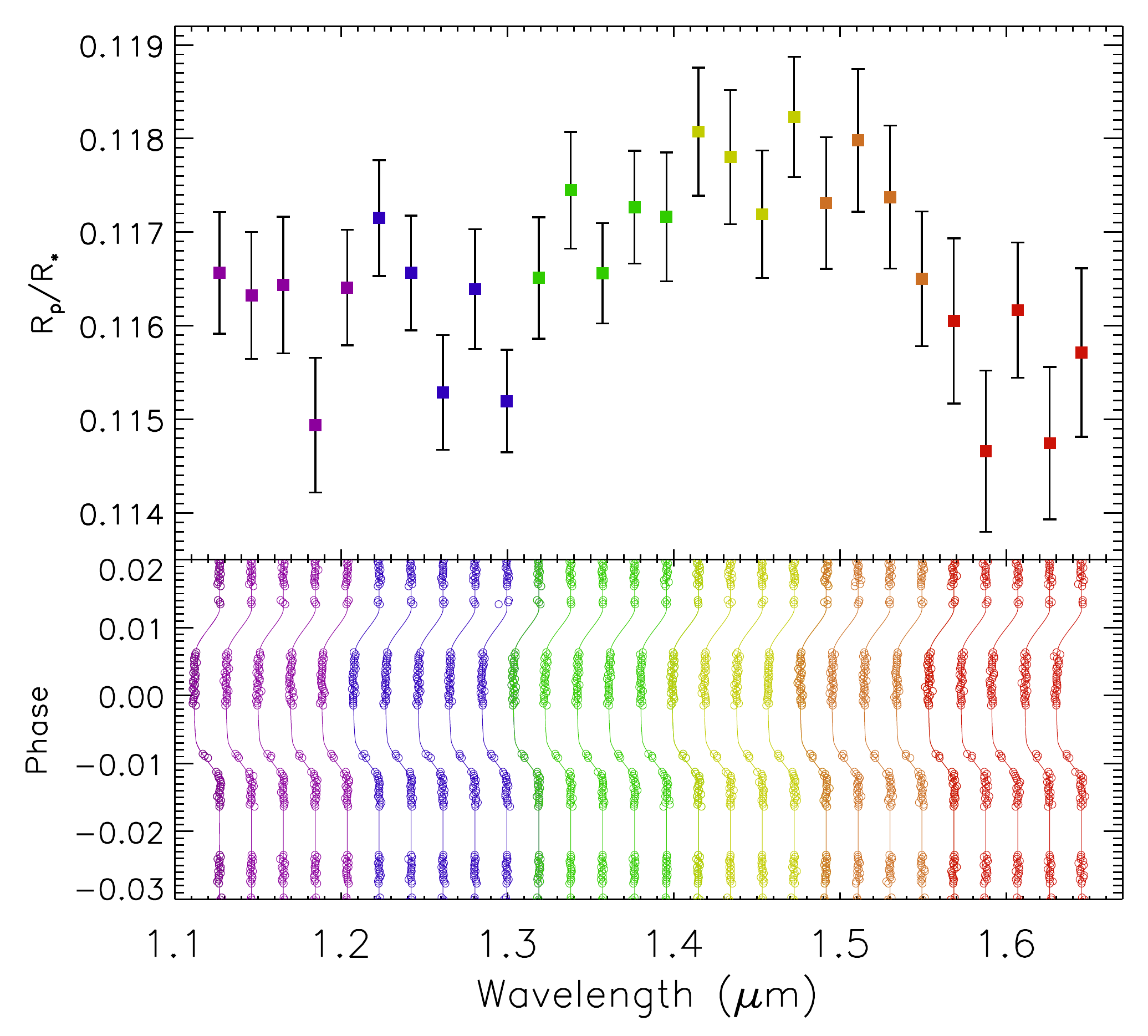}
	\end{center}
\caption{Top: The derived transmission spectrum of HAT-P-1b using differential photometry with individual parameter fitting. Bottom: Spectroscopic light curve for each wavelength bin plotted vertically below the corresponding spectral depth measurement.  The colours are used to guide the eye such that each R$_{p}$/R$_{*}$ measurement can be more easily matched with the corresponding light curve.}
\label{fig:spectrum}
\end{figure*}
	
\begin{table}
\centering	
\caption{Table of constrained system parameters and errors (from \citealt{nikolov2013}).}
\begin{tabular}{ccc}
\hline
\hline
Parameter & Value & Uncertainties \\
\hline
Inclination ($^{\circ}$) & 85.677 &  0.061 \\
Period (days) & 4.46529974 & 0.00000055 \\
a/R$_{*}$ & 9.910 & 0.079 \\
Center of transit time (JD) & 2456114.345307 & 0.00018 \\
\hline
\end{tabular}
\end{table}

\subsection{White light curve fits}

	\indent{} Prior to evaluating the transmission spectrum (from transit light curves in small spectral bins), we analysed the light curves summed over the entire wavelength range. The white light curve was used to improve the general system parameters and quantitatively investigate any instrumental systematics.  \\
	\indent{}  Systematics in the data that effect both the target and reference star are partially removed by performing differential photometry, dividing target-star flux by the reference-star flux (see Fig. \ref{fig:whitend} for a comparison of the raw white light curves)  reducing the residual scatter by a factor of three. Further systematics present in the data, shown in the differential light curve of Fig. \ref{fig:whitend}, display clear orbit-to-orbit trends of increasing flux within each HST orbit in the raw light curve, which we attribute to a ``breathing effect", caused by the thermal expansion and contraction of HST during its orbit. We fit for this similarly to Brown et al. (2001) and Sing et al. (2011), using a 7th order polynomial fit versus HST orbital phase. To avoid over-fitting  the model as a result of adding parameters, we calculated the Bayesian Information Criterion (BIC) that adds a penalty term for the number of parameters in the model, such that the significance of each new parameter can be estimated. To account for breathing systematics in the light curve, while avoiding over fitting, corrections were applied for a general slope over the entire light curve (a correction over the HST visit) as well as a 7th order polynomial in HST phase (a correction per HST orbit). No further trends, such as the spectral trace position and timing of the central HST orbital phase, were found to significantly improve the white light curve fits, we therefore adopt these methods for our final white light fits (Fig. \ref{fig:white}). We note a significant reduction, up to 65\%, in the parameters computed for the HST ``breathing effect" between single target and differential photometry showing that the ability to perform differential photometry is an important aspect of this analysis. We find a decrease in the white light curve residuals from a standard deviation of 400ppm to 160ppm, placing a meaningful number on the reference star as a calibrator.  Telescope systematic errors affect the science and calibrator stars in the same way to a precision of one part per 2400; we address the residual systematics, 3.2 times larger than the photon noise in the case of these observations, using individual parameter analysis.
	
	\indent{} Throughout our analysis, we implemented a Levenberg-Marquardt least-squares minimisation algorithm (L-M) to determine the best-fit parameters for both the planetary system and any systematics inherent in the data. This is done by using the MPFIT IDL routine by \citet{markwardt2009}. \\
	\indent{} To corroborate these results, we also applied a Markov-chain Monte-Carlo (MCMC) data analysis (\citealt{eastman2012a}). While the L-M computes the best fit $\chi^2$ value of the parameters by estimating the parameter errors from the covariance matrix calculated using numerical derivatives, the MCMC computes the maximum likelihood of the parameter fit given a prior value and evaluates the posterior probability distribution for each parameter of the model. The MCMC routine uses a simplified quadratic limb-darkening model described by parameters allowed to vary within the Kurucz grid of stellar spectra as a function of emergent angle. EXOFAST (a fast exoplanetary fitting suite in IDL) also uses the stellar mass-radius relation of \citet{torres2008} to constrain the stellar parameters, compared to fixed non-linear limb-darkening parameters used in the L-M with unconstrained stellar parameters. 
	MCMC can be more robust against finding local minima when searching the parameter space, where the L-M may get trapped.
	\\	
	Each method produces similar results within the errors with the main small differences arising primarily from the different limb-darkening fitting procedures. \\
	\indent{} The system parameters and uncertainties for, orbital inclination, orbital period, a/R$_{*}$, and centre of transit time were constrained using a combined MCMC fit with three HST/STIS transit observations, two using G430L and one using G750L, and our WFC3 transit data (see Table 1). 

\begin{figure}
\begin{center}
	\includegraphics[width=8.5cm]{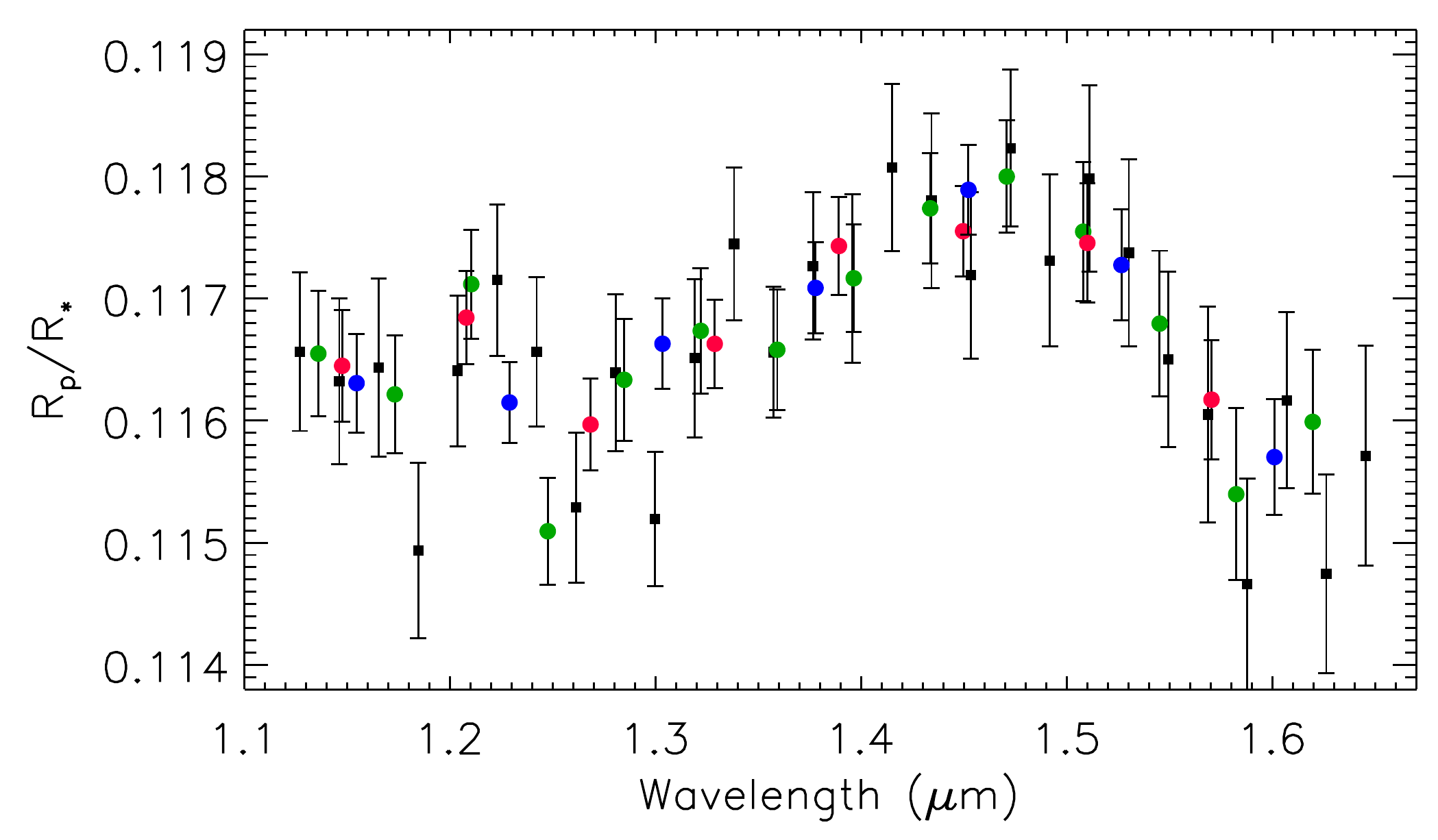}
	\end{center}
\caption{Transmission spectrum of HAT-P-1b, derived using differential photometry with individual-parameter fitting, for $\Delta\lambda =19.2nm$ resolution shown as black squares. Over plotted are the transmission spectra for a range of different wavelength resolution bins: $\Delta\lambda =37.2nm$ in green; $\Delta\lambda =60.4nm$ in pink; and $\Delta\lambda =74.4nm$ in blue.}
\label{fig:binsizes}
\end{figure}

		\indent{} The initial starting values for planetary and system parameters were taken from \citet{butler2006},  \citet{johnson2008}, and \citet{torres2008}. The best fit  light curve for the WFC3 transit along with the uncertainties associated with the computation were determined using MPFIT giving a final white light radius ratio of R$_{P}$/R$_{*}$ = 0.11709$\pm$0.00038 (see Fig. \ref{fig:white}). 
	
\begin{figure}
\begin{center}
	\includegraphics[width=9cm]{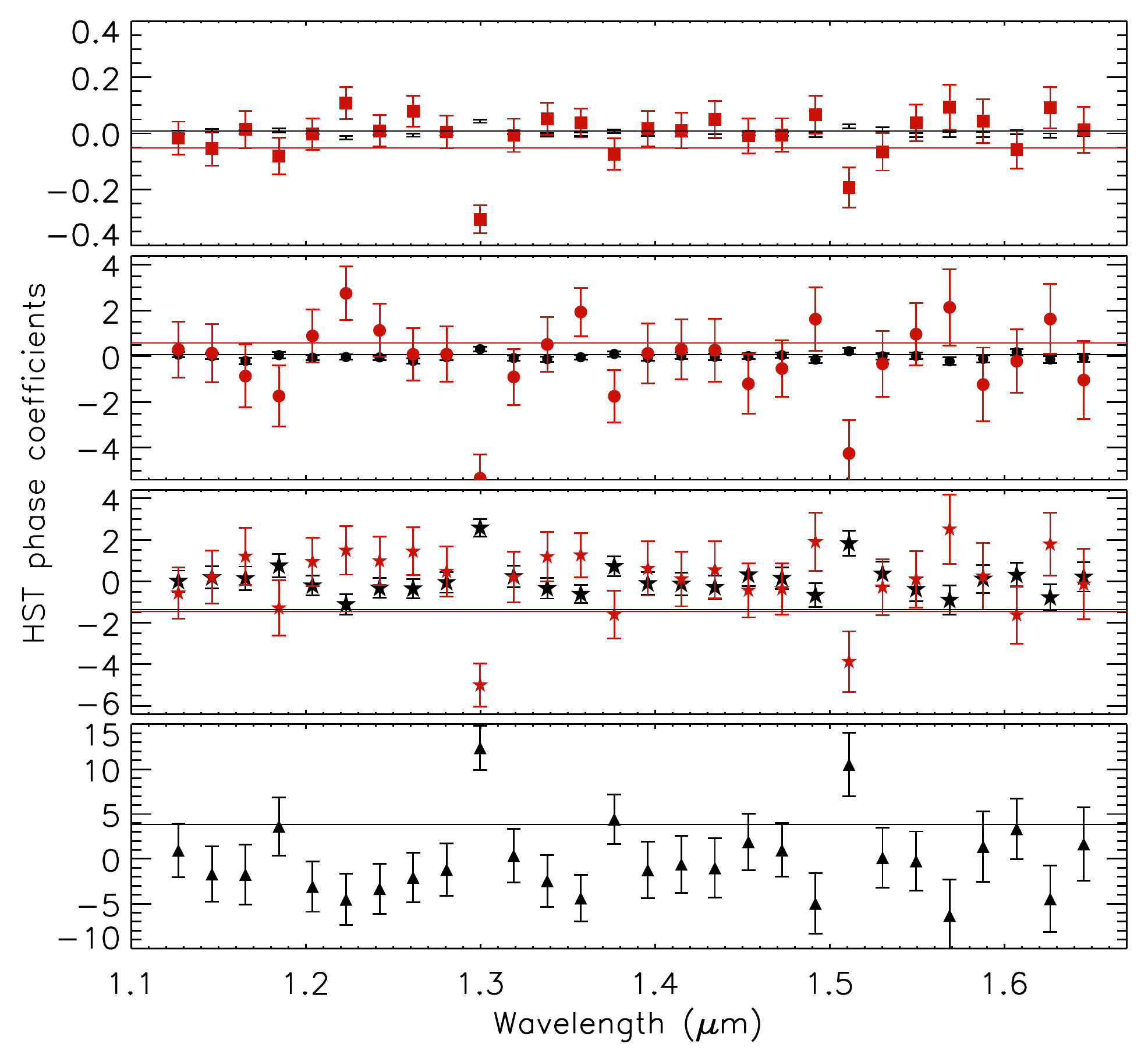}
	\end{center}
\caption{HST phase coefficients for each of the spectroscopic bins using differential photometry individual parameter fitting. Top: The 1st (black), 2nd order coefficients (red, squares). Middle-Top: The 3rd (black-circles) and 4th (red-circles) order coefficients showing a near zero variation over each wavelength bin. Middle-bottom: The 5th (black-stars) and 6th order coefficient (red-stars) Bottom: The 7th order HST phase coefficient for each bin. Note the y axis scale for each plot with the corresponding white light coefficient marked as a solid line. }
\label{fig:hubble}
\end{figure}
\begin{figure}
\begin{center}
	\includegraphics[width=8.5cm]{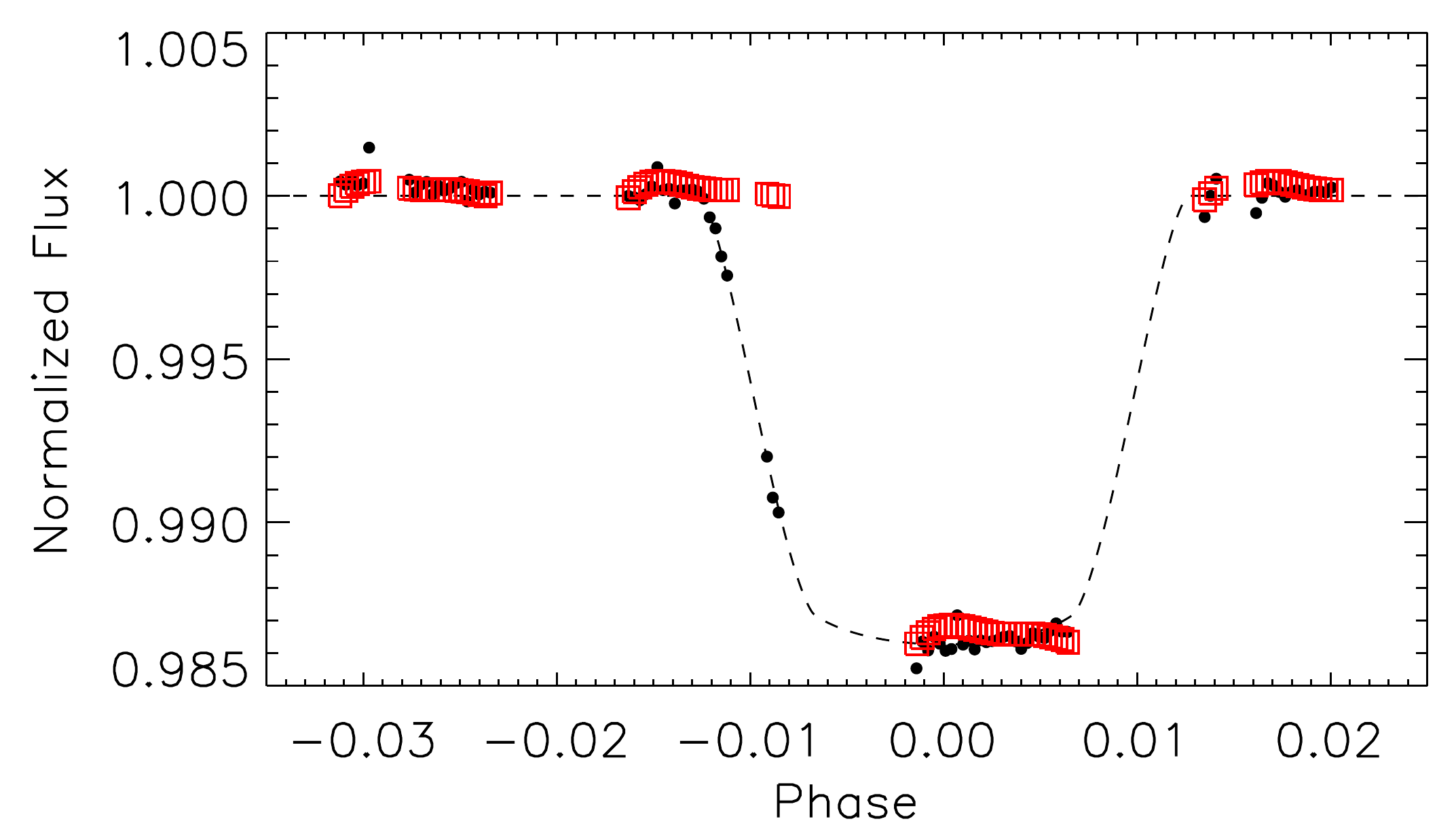}
	\end{center}
\caption{Raw white light curve with the breathing correction function over-plotted as open squares (red) to show the fit to the orbit-to-orbit trends evident in the data corresponding to the 7th order parameter.  }
\label{fig:hstphase}
\end{figure}

	We also fit the white light curve for single target photometry as well as differential photometry as shown in Fig. \ref{fig:white}.  Without differential photometry there are systematics in the data that increase the errors and the deviation from the mean as shown by the residual plot at the bottom of Fig. \ref{fig:white}, which shows that the differential photometry reduces the scatter in the residuals by a factor of three. For both light curves the red noise, defined as the noise correlated with time ($\sigma_r$), is estimated at each time-averaged bin of the light curve containing $N$ points following \citet{pont2006},
	
	 \begin{equation} \label{rednoise}
	 \sigma_{N} = \sqrt{ \frac{\sigma_{w}^{2}}{N} +  \sigma_{r}^{2}}
	 \end{equation}
	where $\sigma_w$ is the white uncorrelated noise and $\sigma_{N}$ is the photon noise. For our best fit light curve, we find $\sigma_{w} = 1.49\times10^{-4}$, with $\sigma_{r} = 4.97\times10^{-5}$ using a bin size of N=10 (see Fig. \ref{fig:white}) with a photon noise level of $6.8\times10^{-5}$.	

	\indent{} Another method used to empirically correct for repeating systematics between orbits is the divide-oot routine developed by \cite{berta2012}. Divide-oot uses the out-of-transit orbits to compute a weighted average of the flux evaluated at each exposure within an orbit and divides the in-transit orbits by the template created. This requires each of the in-transit exposures to be equally spaced in time with the out-of-transit exposures being used to correct them, so that each corresponding image has the same HST phase so that additional systematic effects are not introduced. Due to this constraint, we were unable to perform the out-of-transit method as both the in-transit and out-of-transit orbits contain a different number of exposures with varied spacing between exposures. 
The divide-oot method relies on the cancellation of common-mode systematic errors by operating only on the data themselves using simple linear procedures, relying on trends to be similar in the time domain. A somewhat similar technique was adopted by \citet{deming2013} for their analysis of WFC3 data relying on common trends in the wavelength domain. In Sec. 3.4.1 we adopt a similar method of subtracting white-light residuals from each spectroscopic bin to corroborate our results from individual parameter analysis.

\begin{table}
	\centering
	\caption{Transmission spectrum and limb-darkening coefficients for HAT-P-1b from WFC3/G141 using differential photometry and with common-mode removal of systematic errors (see Fig. \ref{fig:spectrum}).}
\begin{tabular}{cccccc}
	\hline
	\hline
	$\lambda$ & R$_p$/R$_*$ & c$_{1}$ & c$_{2}$ & c$_{3}$ & c$_{4}$\\
	($\mu$m) & ~ & ~&~&~&~ \\
	\hline        	                                                                
1.1269 & 0.11656 $\pm$0.00065 & 0.7301 & -0.4003 & 0.3529 & -0.1200 \\
1.1461 & 0.11632 $\pm$0.00068 & 0.7271 & -0.3993 & 0.3497 & -0.1186 \\
1.1653 & 0.11643 $\pm$0.00073 & 0.7253 & -0.4005 & 0.3399 & -0.1133 \\
1.1845 & 0.11493 $\pm$0.00072 & 0.7192 & -0.3703 & 0.3011 & -0.0981 \\
1.2037 & 0.11640 $\pm$0.00062 & 0.7157 & -0.3677 & 0.2922 & -0.0939 \\
1.2229 & 0.11715 $\pm$0.00062 & 0.7273 & -0.3759 & 0.2904 & -0.0917 \\
1.2421 & 0.11656 $\pm$0.00061 & 0.7315 & -0.3769 & 0.2802 & -0.0859 \\
1.2613 & 0.11528 $\pm$0.00061 & 0.7349 & -0.3673 & 0.2553 & -0.0743 \\
1.2805 & 0.11639 $\pm$0.00064 & 0.7639 & -0.4002 & 0.2308 & -0.0562 \\
1.2997 & 0.11519 $\pm$0.00055 & 0.7470 & -0.3724 & 0.2322 & -0.0606 \\
1.3189 & 0.11651 $\pm$0.00065 & 0.7482 & -0.3768 & 0.2326 & -0.0599 \\
1.3381 & 0.11744 $\pm$0.00062 & 0.7560 & -0.3824 & 0.2219 & -0.0525 \\
1.3573 & 0.11656 $\pm$0.00054 & 0.7710 & -0.4064 & 0.2325 & -0.0538 \\
1.3765 & 0.11726 $\pm$0.00060 & 0.7885 & -0.4378 & 0.2473 & -0.0553 \\
1.3957 & 0.11716 $\pm$0.00069 & 0.8061 & -0.4666 & 0.2591 & -0.0558 \\
1.4149 & 0.11807 $\pm$0.00068 & 0.8292 & -0.5034 & 0.2796 & -0.0598 \\
1.4341 & 0.11780 $\pm$0.00072 & 0.8522 & -0.5623 & 0.3265 & -0.0735 \\
1.4533 & 0.11719 $\pm$0.00068 & 0.8706 & -0.5906 & 0.3363 & -0.0729 \\
1.4725 & 0.11823 $\pm$0.00064 & 0.8915 & -0.6199 & 0.3506 & -0.0747 \\
1.4917 & 0.11731 $\pm$0.00070 & 0.9156 & -0.6854 & 0.4058 & -0.0917 \\
1.5109 & 0.11798 $\pm$0.00076 & 0.9470 & -0.7560 & 0.4641 & -0.1095 \\
1.5301 & 0.11737 $\pm$0.00076 & 0.9788 & -0.8295 & 0.5260 & -0.1288 \\
1.5493 & 0.11650 $\pm$0.00072 & 0.9714 & -0.8486 & 0.5619 & -0.1451 \\
1.5685 & 0.11605 $\pm$0.00088 & 0.9875 & -0.9154 & 0.6342 & -0.1719 \\
1.5877 & 0.11466 $\pm$0.00086 & 1.0501 & -1.0948 & 0.8137 & -0.2357 \\
1.6069 & 0.11616 $\pm$0.00072 & 1.1217 & -1.2570 & 0.9557 & -0.2820 \\
1.6261 & 0.11474 $\pm$0.00081 & 1.1263 & -1.2696 & 0.9679 & -0.2861 \\
1.6453 & 0.11571 $\pm$0.00090 & 1.0649 & -1.1631 & 0.8794 & -0.2593 \\
	\hline
\end{tabular}
\end{table}

	\begin{figure}
\begin{center}
	\includegraphics[width=9cm]{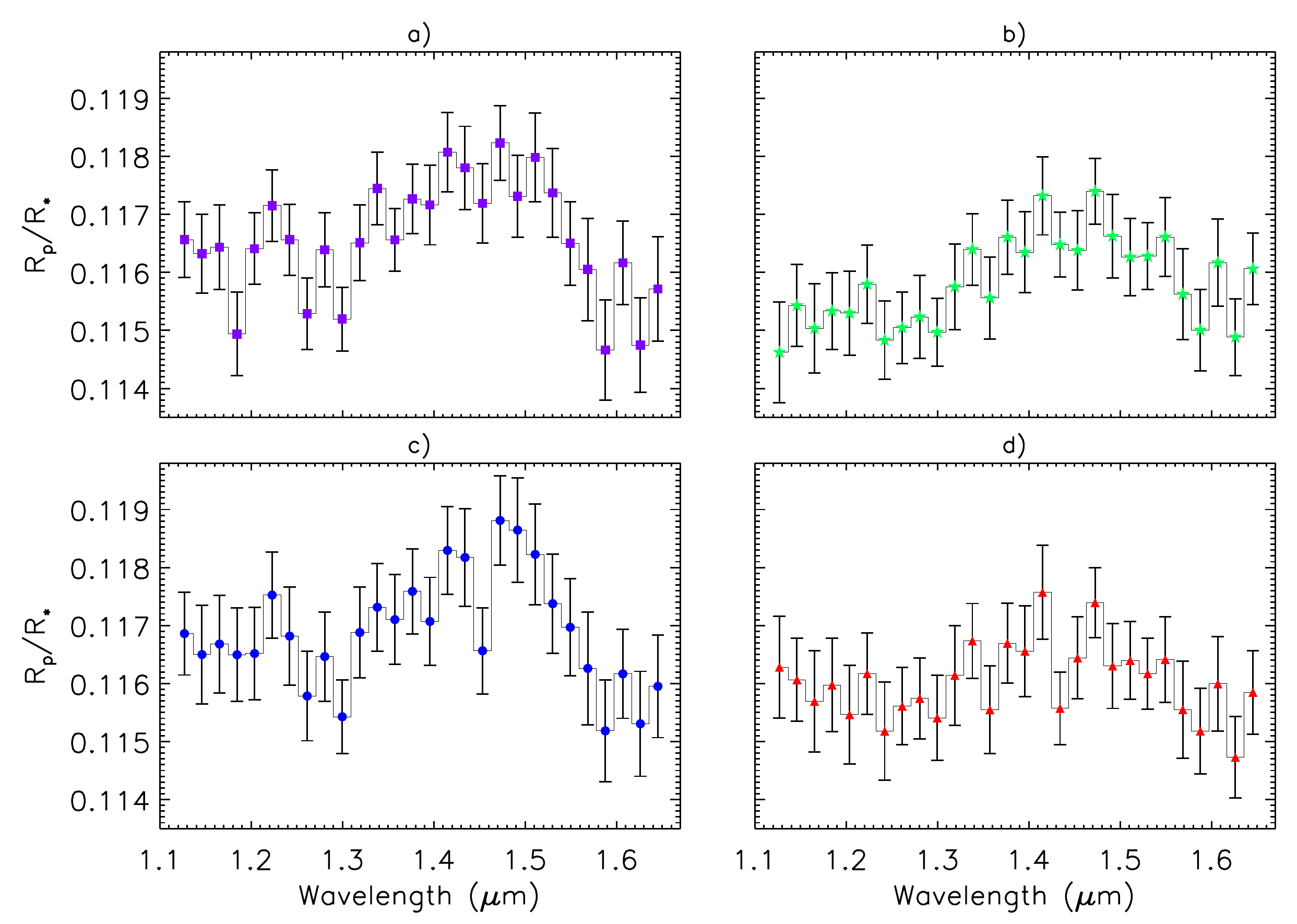}
	\end{center}
\caption{a) Transmission spectrum of HAT-P-1b for differential photometry individual parameter fitting. b) Single target photometry individual parameter fitting. c) Differential photometry with common mode fitting. d) Single target photometry with common mode fitting. While each spectra show a common spectral shape the method used for figure a) has lower red noise  and residual scatter for each spectroscopic bin and is therefore adopted transmission spectrum for further analysis. }
\label{fig:alltranspec}
\end{figure}

\begin{figure}
\begin{center}
	\includegraphics[width=8.5cm]{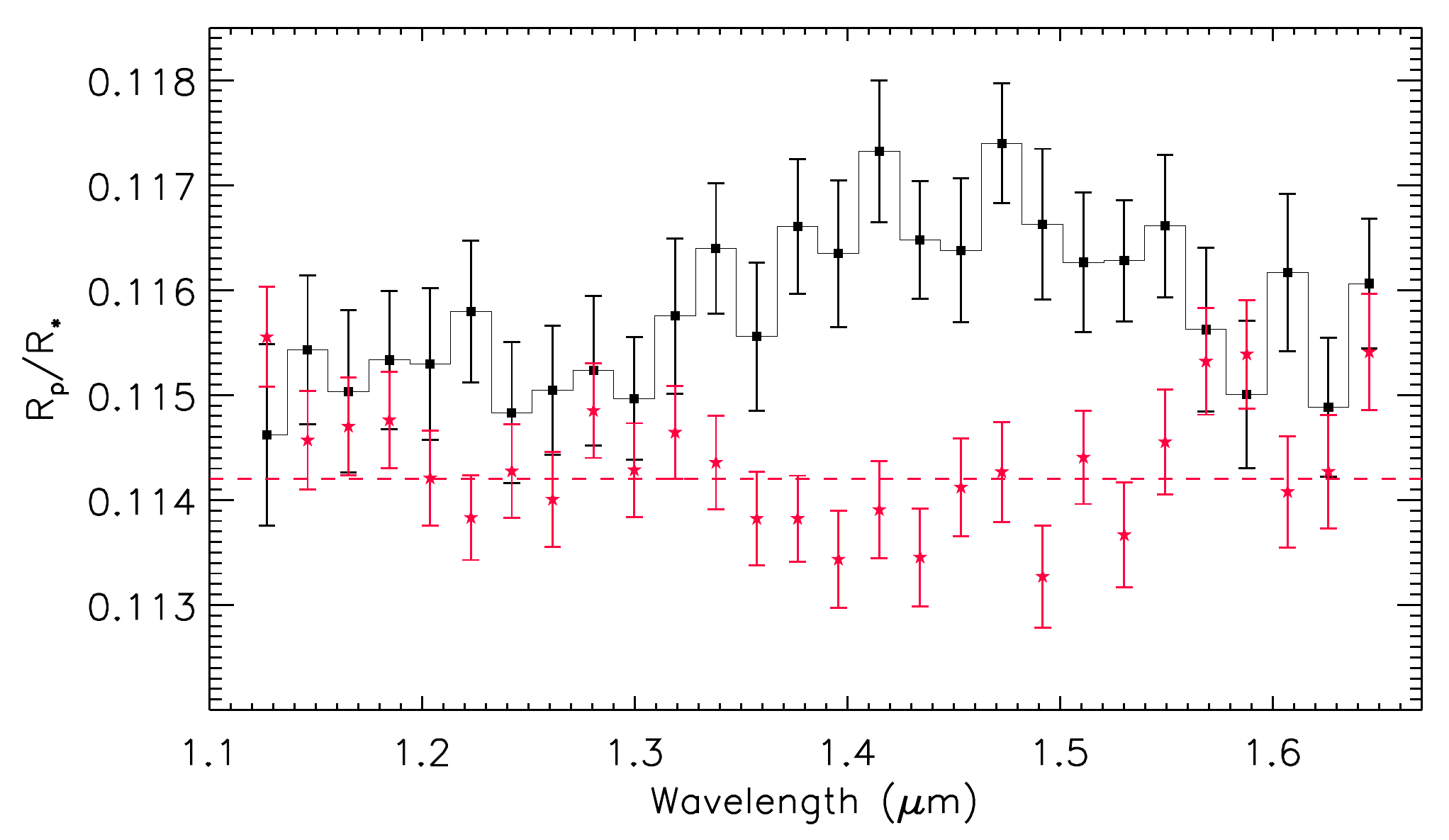}
	\end{center}
\caption{ Plotted in red stars is a transmission spectrum for the reference star computed after injecting a transit of constant depth (represented by the dashed red line) into the light curve. The black squares show the transmission spectrum of HAT-P-1b using single target photometry and individual parameter systematic fitting. The `transit spectrum' of the reference star is rather flat, and does not show the water absorption spectral shape.}
\label{fig:inject}
\end{figure}

\begin{figure*} 
\begin{center}
	\includegraphics[width=18cm]{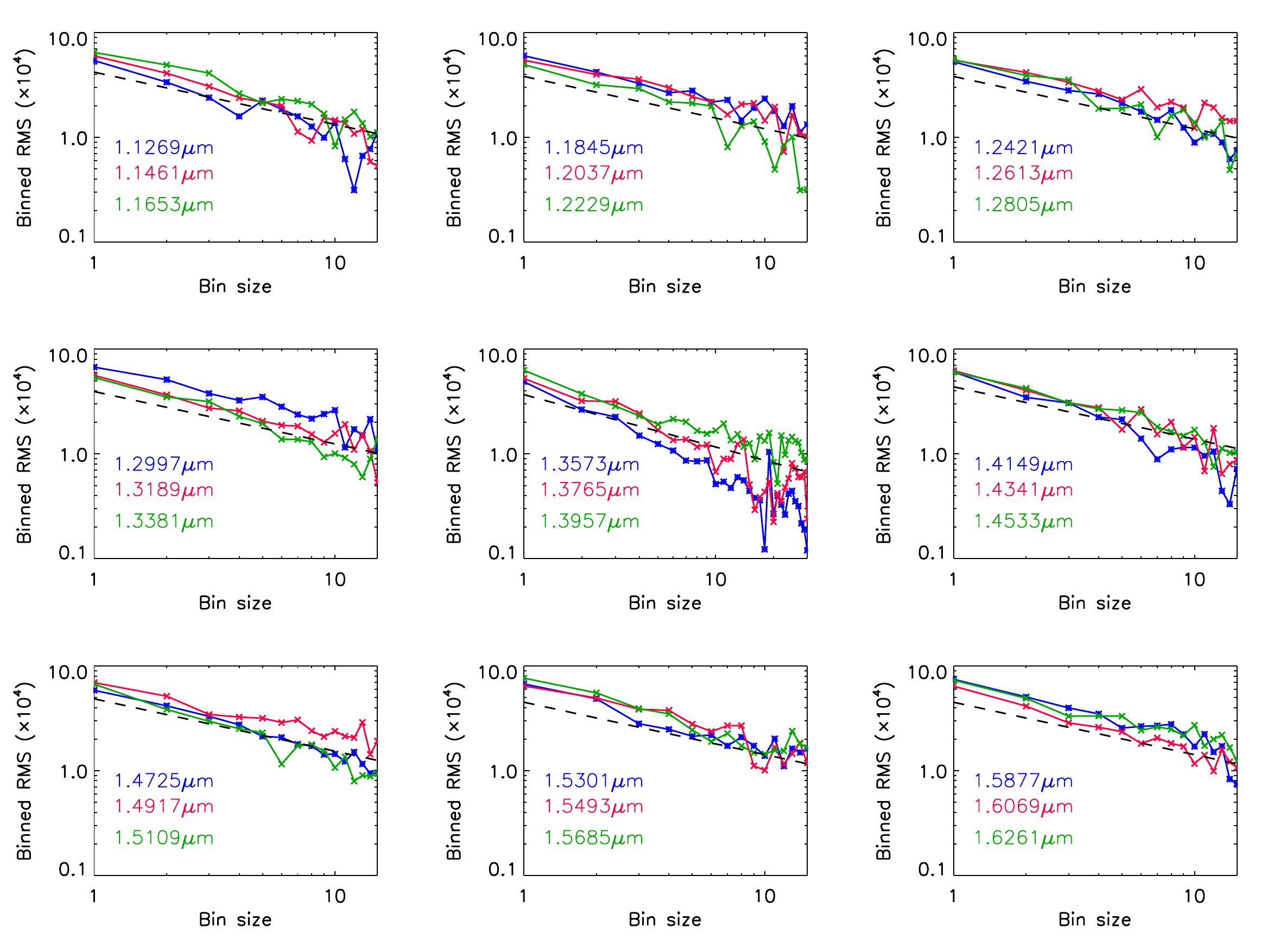}
	\end{center}
\caption{ Binned root mean square of the residuals for each spectroscopic bin (red, blue, and green) plotted against the photon noise for the central wavelength channel of each plot (black). The residuals are calculated using differential photometry with individual parameter analysis. }
\label{fig:rednoiseall}
\end{figure*}

\subsection{Spectroscopic light curve fits}

		\indent{} In order to understand and monitor the significance of each potentially common-mode systematic inherent in the WFC3 data we determine a fit for each separate parameter as well as applying a general common mode analysis using the white light residuals. We construct multi-wavelength spectroscopic light curves by binning the extracted spectra into 28 channels that are $\sim$4 pixels wide ($\Delta$$\lambda$ = 0.0192\,$\mu$m) from 1.1173$\mu$m to 1.6549\,$\mu$m, which is close to the resolution of the G141 grism. To measure the transmission spectrum of HAT-P-1b, we conducted individual parameter fitting to each of the 28 light curves with a model in which R$_p$/R$_*$, a baseline flux, and a 7th order polynomial as a function of HST orbital phase are allowed to vary, and with the orbital inclination, orbital period, a/R$_*$,  and the centre of transit time fixed from the white light curve fitting. To avoid over fitting the data, and to determine the consistency of the systematic model used, we computed the BIC number for each spectroscopic bin. The systematic model with the lowest BIC was found to be consistent with that for the white light curve with little significant variation in the computed BIC number between each of the spectroscopic bins. For limb-darkening coefficients we again used the 3D models, fixed for each spectroscopic bin as listed in Table 2. 

	\indent{} Similar to the white light curve, the 7th order HST phase correction is used to account for breathing systematics. The fitted $R_{p}/R_{*}$ for each spectroscopic bin are listed in Table 2 along with the corresponding limb darkening parameters. Figure \ref{fig:spectrum} shows the resultant transmission spectrum as well as the light curves for differential photometry  with individual parameter fitting. The binned root mean squared of the residuals for each wavelength-bin can be seen in Fig. \ref{fig:rednoiseall} shown relative to the photon noise of a representative spectral channel.
\begin{figure*} 
\begin{center}
	\includegraphics[width=16cm]{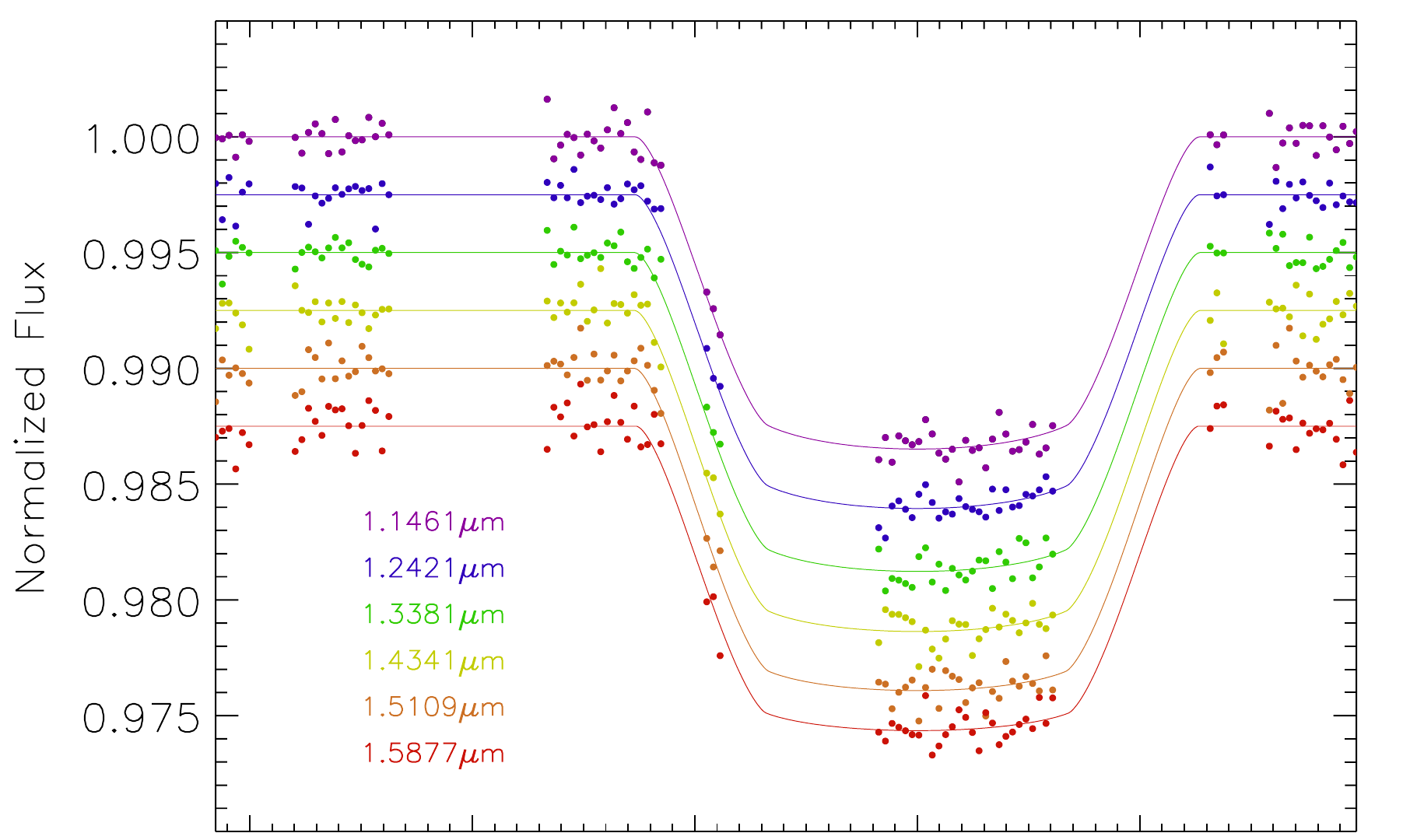}
	\end{center}
\label{fig:lc1}
\end{figure*}

\begin{figure*} 
\begin{center}
	\includegraphics[width=14.8cm]{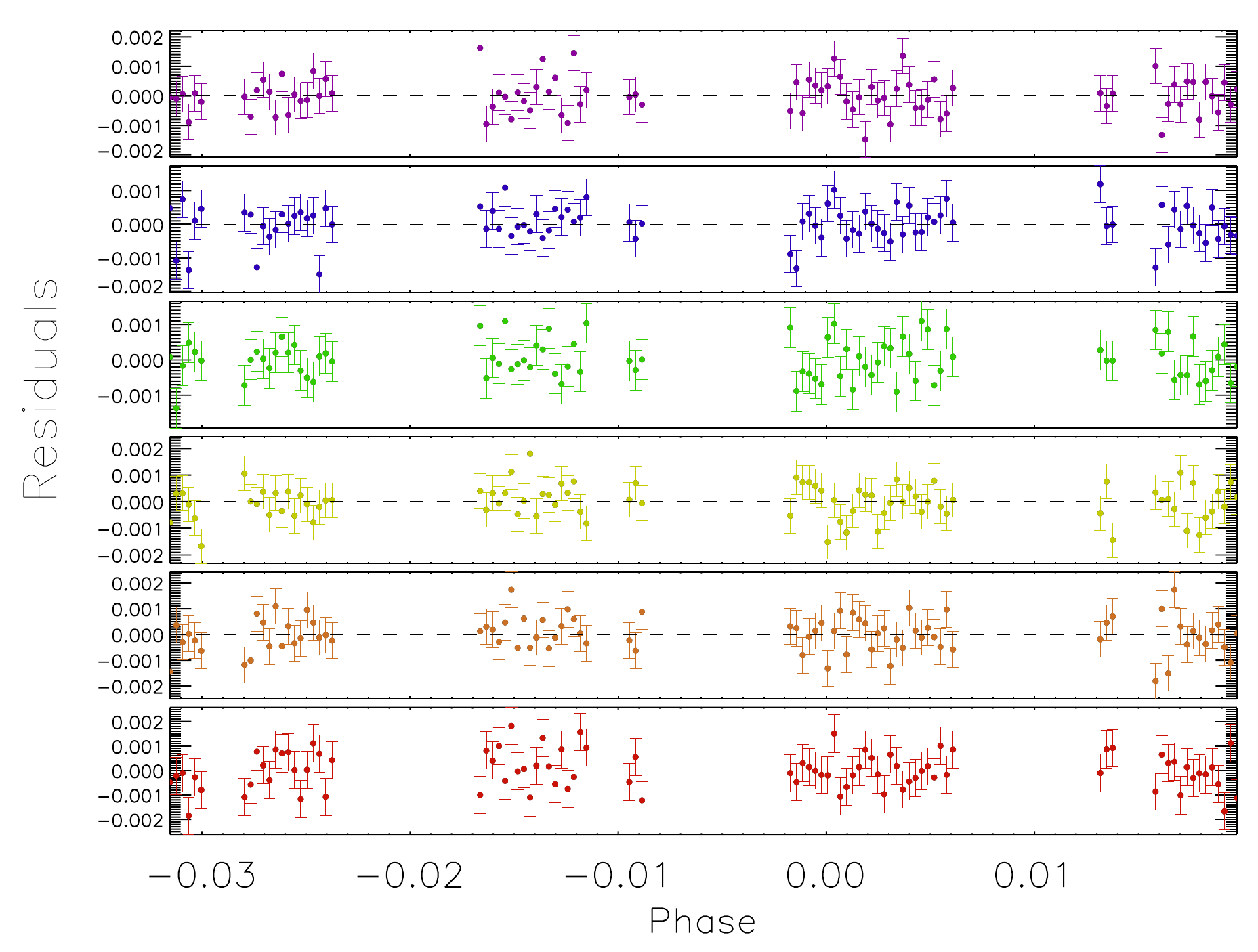}
	\end{center}
\caption{Spectroscopic lightcurves for six different wavelength bins using differential photometry with individual parameter fitting the colours correspond to those used in Fig. \ref{fig:spectrum}.}
\label{fig:speclc}
\end{figure*}
	We derived the transmission spectrum of HAT-P-1b for a wide range of wavelength bin sizes to test the fits and determine the achievable level of precision for the final transmission spectrum (see Fig. \ref{fig:binsizes}). Using differential photometry with individual parameter fitting, we achieved S/N levels of $\sim$1840 per image at a resolution of $\Delta\lambda$ = 19.2\,nm (R$\sim$70). The resulting transmission spectrum consists of 28 bins, each measured to a precision of about one planetary scale height. This demonstrates WFC3's ability to measure the transmission spectrum of exoplanets down to the resolution of the instrument meaning that fine structure in the NIR spectrum of an exoplanetary atmosphere can potentially be measured. 	

\subsubsection{Common mode Systematics}
	WFC3 exhibits common-mode systematic errors across the detector that are predominately not wavelength dependent. Common-mode systematic trends usually do not highly impact the shape of the transmission spectrum, as each spectroscopic light curve bin is similarly effected, and the relative planetary radius information is preserved. The common-mode trends can be seen in Fig. \ref{fig:hubble}, which shows that most of the wavelength bins have a common HST coefficient within 1$\sigma$ of the computed white-light coefficients up to the 7th order. Figure \ref{fig:hstphase} shows this breathing correction for the white light curve over plotted on the raw data showing the correction of a repeating trend in the data. In addition to the breathing systematic, there is also a non-repeating trend evident in the white light residuals (see Fig. \ref{fig:white} and \ref{fig:whitend}), specifically in the 3rd and 4th orbits, that is present in each wavelength band. 
	\begin{table}
\centering	
\caption{ Quantitative analysis of each analysis method used to compute the transmission spectra displayed in Fig. 9. This shows the significant decrease in red noise computed for differential photometry with individual parameter analysis with an additional decrease in the standard deviation of the residuals when compared to common-mode removal.}
\begin{tabular}{p{2.5cm}p{1cm}p{1cm}p{1cm}p{1cm}}
\hline
\hline
Figure 9 & a) & b) & c)  & d) \\
\hline
Standard deviation of the residuals & 0.00062 & 0.00059 & 0.00076 & 0.00071 \\
Red noise & 0.00001 & 0.00008 & 0.00016 & 0.00014 \\
($\sim$8 minute bins)  &~&~&~&~\\
$\sigma_{N}$  & 0.00019 & 0.00020 & 0.00029 & 0.00026 \\
($\sim$8 minute bins)  &~&~&~&~\\
BIC & 131 & 133 & 142 & 150 \\
\hline
\end{tabular}
\end{table}

\begin{figure}
\begin{center}
	\includegraphics[width=9cm]{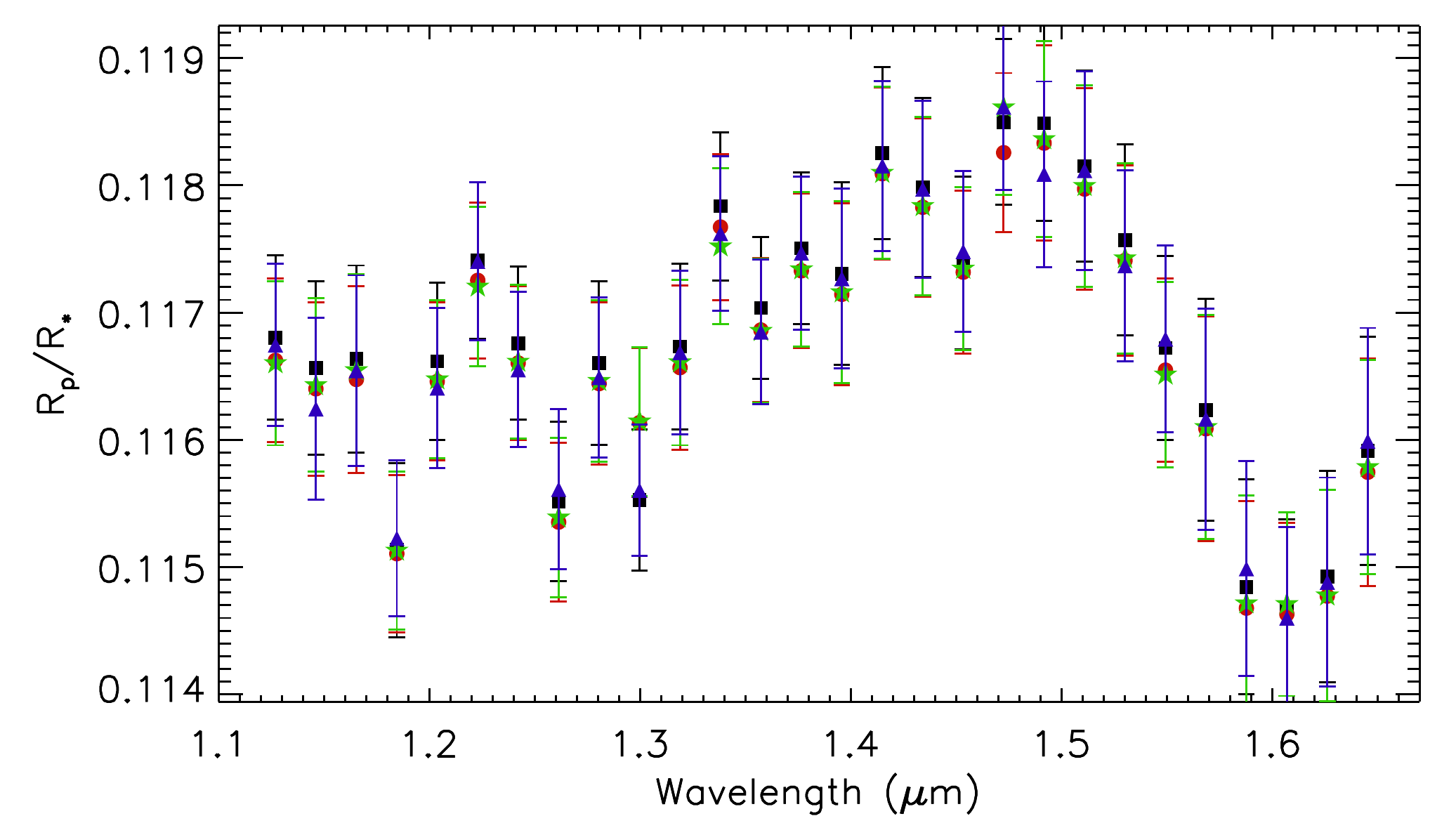}
	\end{center}
\caption{HAT-P-1b transmission spectrum computed for differential photometry with individual parameter analysis for four different systematic models. 7th order polynomial in HST orbital phase (black squares), 6th order polynomial in HST orbital phase (red circles), 5th order polynomial in HST orbital phase (green stars), 4th order polynomial in HST orbital phase (blue triangles). }
\label{fig:orders}
\end{figure}

	We therefore calculated the transmission spectrum using four different methods, testing the effects of individual parameter fitting and the cancelation of common-mode systematics using simple linear procedures, for both differential and single-target photometry. The four different methods displayed in Fig. \ref{fig:alltranspec} show a common structure to the transmission spectrum, indicating the significance of the spectral feature despite the assorted analysis techniques regarding differential analysis and common mode removal of the systematic trends.  We choose to quote final values for the transmission spectra from the analysis using differential photometry with individual parameter fitting, as it produces the highest quality light curve. The mean scatter of the residuals for all of the spectral bins is reduced by 10\% from single to differential photometry. In addition a reduction of $\sim$20\%  is seen between common-mode removal and individual parameter analysis. There is also a significant reduction in the red noise from $\sigma_r$=1.4$\times$10$^{-4}$ for differential photometry with common mode removal down to $\sigma_r$=0.1$\times$10$^{-4}$ for differential photometry with individual parameter analysis (see Table 3). In addition, by conducting both differential photometry and individual parameter analysis, we are thus able to better budget for the affects of the dominant thermal-breathing systematic on the transit depths (through the use of the covariance matrix) and to better understand the specific wavelength dependant systematics inherent in the WFC3 data. While small, these can still potentially affect the resultant spectrum obtained. We have therefore adopted the method corresponding to Fig. \ref{fig:alltranspec}a for further analysis and model fitting. We also perform analysis on the transmission spectrum in Fig. \ref{fig:alltranspec}b discussed in Sec. 4.1.1 to corroborate the absorption significance of the water absorption feature. Figure \ref{fig:speclc} shows six of the 28 wavelength channels and their corresponding light curves fitted with differential photometry and individual parameter analysis; the residuals demonstrate that this method efficiently corrects for the apparent common mode trend seen in the white light residuals in Fig. \ref{fig:white}.  We compute the transmission spectrum for differential photometry over a number of systematic models from 4th order polynomial in HST orbital phase to a 7th order polynomial in HST orbital phase adopted for this analysis (see Fig. \ref{fig:orders}). Figure \ref{fig:orders} shows that systematic models fitting for HST orbital phase with a polynomial in the order between 4th and 7th do not change the overall transmission spectrum while the BIC analysis favours a 7th order polynomial fit to the data. As we cannot use the divide-oot routine, there are still some un-modelled systematics in the white light curve data resulting in a precision 2.9 times the photon limit. Though we note that the absolute white light precision per exposure is $\sim$2.3 times better than Berta et al. (2012). With the use of optimised scheduling future observations can potentially take advantage of divide-oot with spatial scanning to increase the white light performance. Our spectroscopic measurements come close to the photon noise limit of the detector with a mean error within 12\% of the photon-limit and a precision of Rp/R* less than 0.0009 per spectral channel similar to that shown by Deming et al. (2013) and \citet{swain2012}. \\	
	
\begin{figure*}
\begin{center}
	\includegraphics[width=16cm]{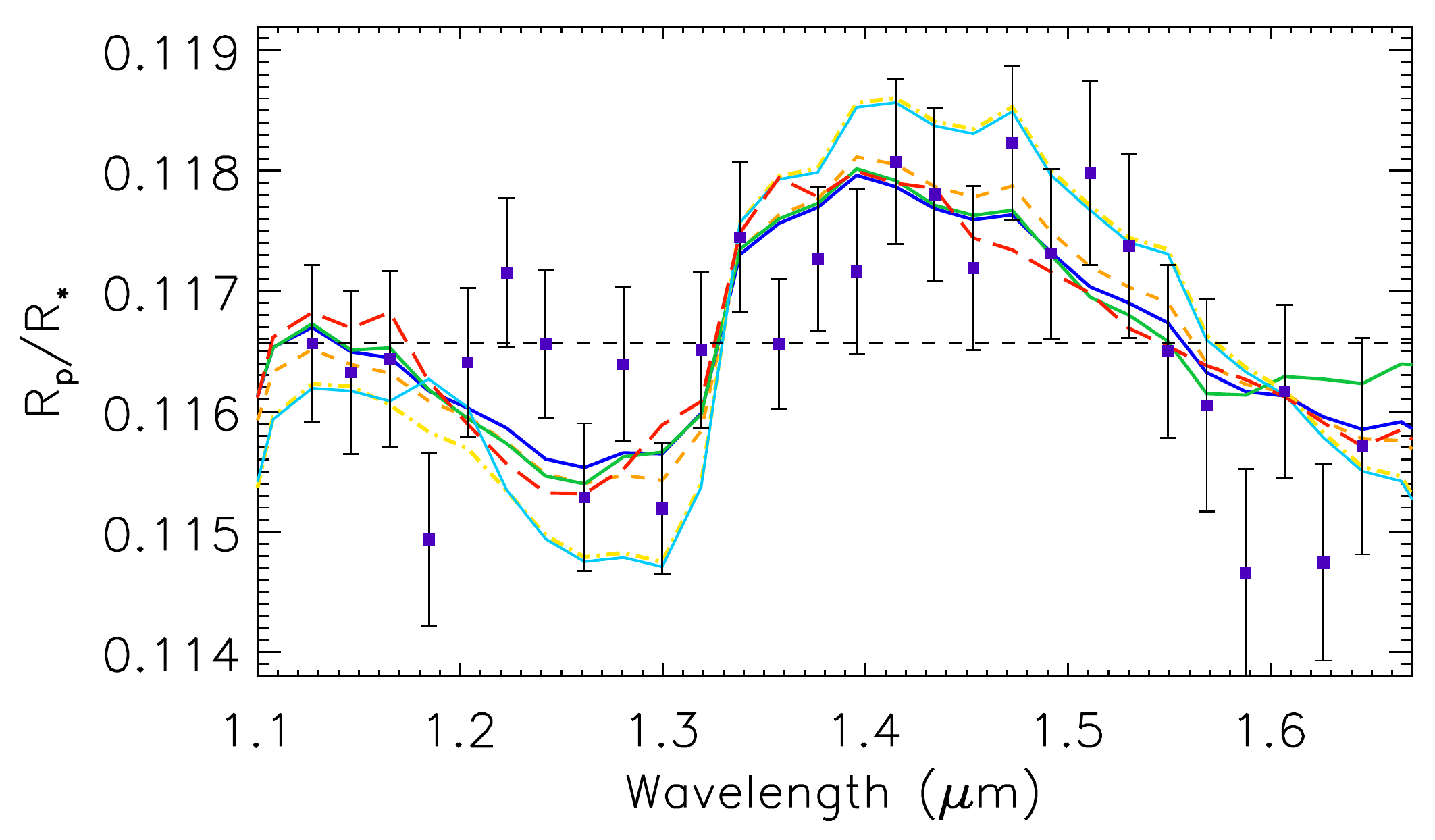}
	\end{center}
\caption{The transmission spectrum of HAT-P-1b, derived using differential photometry with individual parameter fitting (see Fig. 9a). Each theoretical transmission spectrum discussed in section 4.1 is plotted over the data; Orange dashed: hotter dayside-averaged T-P profile model. Dark blue: cooler planetary averaged T-P profile. Red long dashed: dayside model without TiO/VO. Green: isothermal 1000K model. Yellow dot-dash: isothermal 1500K with TiO/VO. Pale blue: isothermal 1500K no TiO/VO.}
\label{fig:model}
\end{figure*}	

			\indent{} Finally, to further characterise systematic effects in� the data that may not have been accounted� for,� we injected a transit of constant depth (R$_{p}$/R$_{*}$=\,0.1142) into the reference star's light curve and computed the transmission spectrum over the same wavelength range with the same bin size. To compute the transmission spectrum, 7th order HST orbital phase corrections were applied and no common-mode systematic removal was conducted. The resultant transmission spectrum shows the wavelength variation in the flux of the reference star using the same exposures used to measure the planetary transit, and can be directly compared to the transmission spectrum �of HAT-P-1b computed using single target photometry and individual parameter (i.e. with 7th order HST orbital phase correction and no common-mode systematic removal) (see Fig. \ref{fig:inject}). \\
	\indent{} As expected, the computed reference star `transit spectrum' is flat, with no water feature observed at 1.4\,$\mu$m. This further demonstrates the� reliability� of  the derived transit spectrum over the whole G141 spectral range.

\begin{figure}
\begin{center}
	\includegraphics[width=8.7cm]{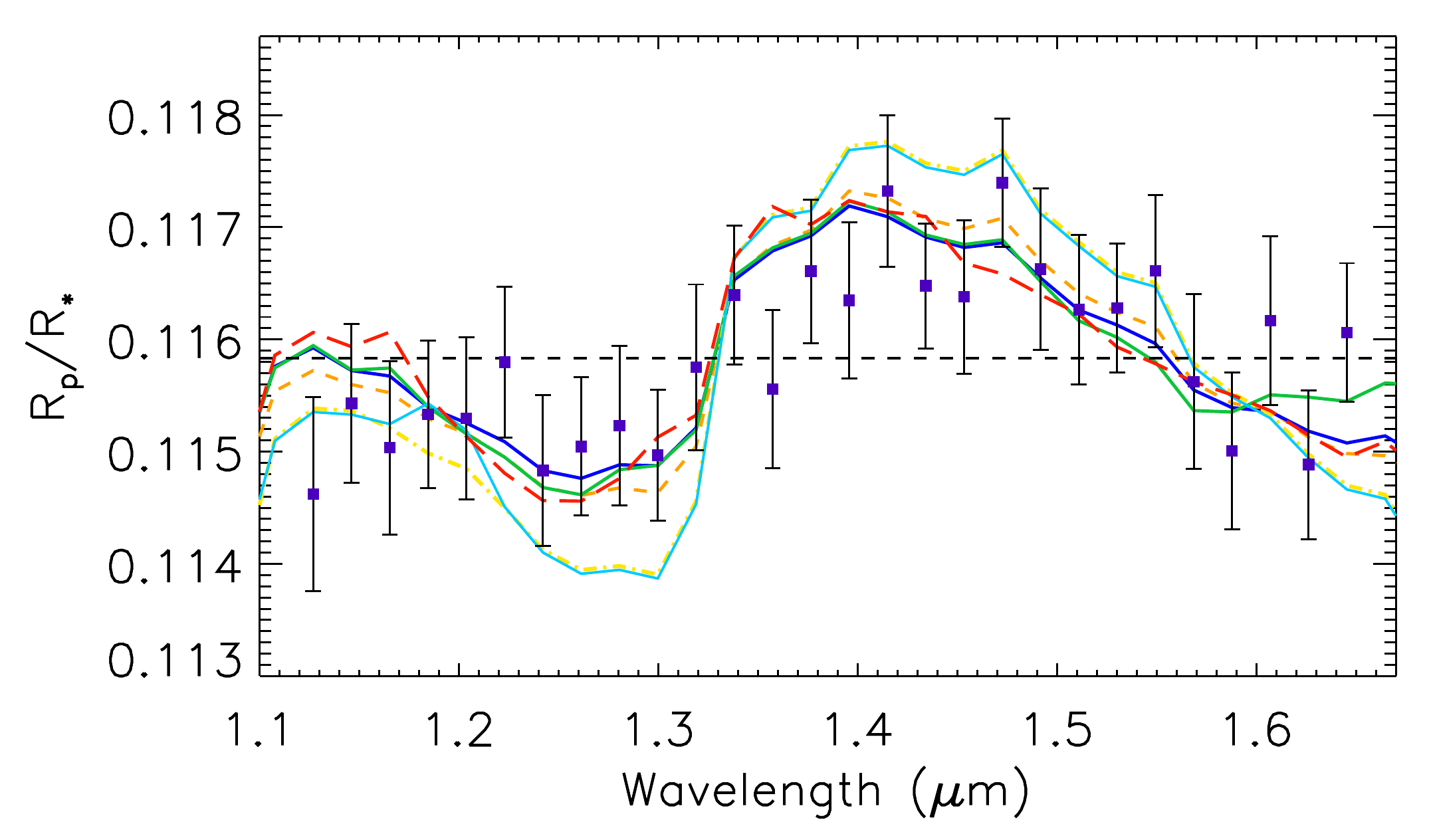}
	\end{center}
\caption{The transmission spectrum of HAT-P-1b, derived using single target photometry with individual parameter fitting (see Fig. 9b). Each theoretical transmission spectrum discussed in section 4.1 is plotted over the data; Orange dashed: hotter dayside-averaged T-P profile model. Dark blue: cooler planetary averaged T-P profile. Red long dashed: dayside model without TiO/VO. Green: isothermal 1000K model. Yellow dot-dash: isothermal 1500K with TiO/VO. Pale blue: isothermal 1500K no TiO/VO.}
\label{fig:modelb}
\end{figure}

%
\section{Discussion}
	\indent{} The transmission spectrum of HAT-P-1b around 1.4\,$\mu$m is presented in Fig. \ref{fig:spectrum}. We compare the transmission spectrum to theoretical atmospheric models of HAT-P-1b based on the models from \citet{fortney2010} and \citet{burrows2013}.

	\indent{} Over the observed wavelength range sampled by the WFC3 G141 grism, the strongest atmospheric feature expected is water absorption band with a characteristic bandhead at 1.4\,$\mu$m. In most lower atmosphere models of hot Jupiters H$_{2}$O is well mixed throughout the atmosphere, and most of the features between 0.7 and 2.5\,$\mu$m come from the H$_{2}$O vibration-rotation bands (\citealt{brown2001a}). These features are difficult to measure with ground-based telescopes due to confusion with water vapour signatures from the Earth's atmosphere. Space-based observations are therefore essential to probe such spectral regions in exoplanetary atmospheric studies. \\
	\indent{} To help interpret the size of the spectral features seen in the transmission spectrum, we determine the scale height of the atmosphere that defines potential spectral features. The scale height (H) is the altitude range over which the atmospheric pressure decreases by a factor of \emph{e},
	\begin{equation} \label{scaleheight}
	 H = \frac{k_{B}T}{\mu_m m_H g}, 
	\end{equation}
 where, $k_{B}$ is the Boltzman constant, $T$ is the estimated atmospheric temperature, $m_H$ is the mass of hydrogen atom, $\mu_m$ is the mean molecular weight of the atmosphere, and $g$ is the surface gravity. The scale height of HAT-P-1b is approximately 500\,km for a H, He atmosphere at T = 1200\,K, which corresponds to transit depths of $\sim0.017\,\%$ or 0.00062 R$_{p}$/R$_{*}$. If water is to be observed in the NIR transmission spectrum of HAT-P-1b then the size of absorption features should be approximately two scale heights or more in size, which is well within the accuracy of these observations (see Fig. \ref{fig:model} and Fig. \ref{fig:model}). 


\subsection{Atmospheric models for HAT-P-1b}
\indent{} We compared the derived transit spectrum of HAT-P-1b to two different suites of theoretical atmospheric models for the transmission spectra, one set of models based on the formalism of  \citet{burrows2010} and the other set based on the models by Fortney et al. (2008; 2010). The pre-calculated models were compared to the data in a $\chi^2$ test, with the base planetary radius as the only free parameter to simply adjust the overall altitude normalisation of the model spectrum. As no interaction is made directly with the model parameters  when making a comparison, such as fitting for the abundance of TiO/VO, H$_{2}$O, or T-P profile, the degrees of freedom for the $\chi^2$ test does not change between models. This analysis aims to distinguish between a number of the different assumptions used in current models, and to identify any expected spectral features rather than to perform spectral retrieval. The transmission spectrum is therefore compared to previously published models of  \citet{burrows2010} and Fortney et al. (2008; 2010) calculated for the radius, gravity, orbital distance, and stellar properties of the HAT-P-1 system. This was done for both isothermal models as well as planetary specific models.

\begin{figure}
\begin{center}
	\includegraphics[width=8.5cm]{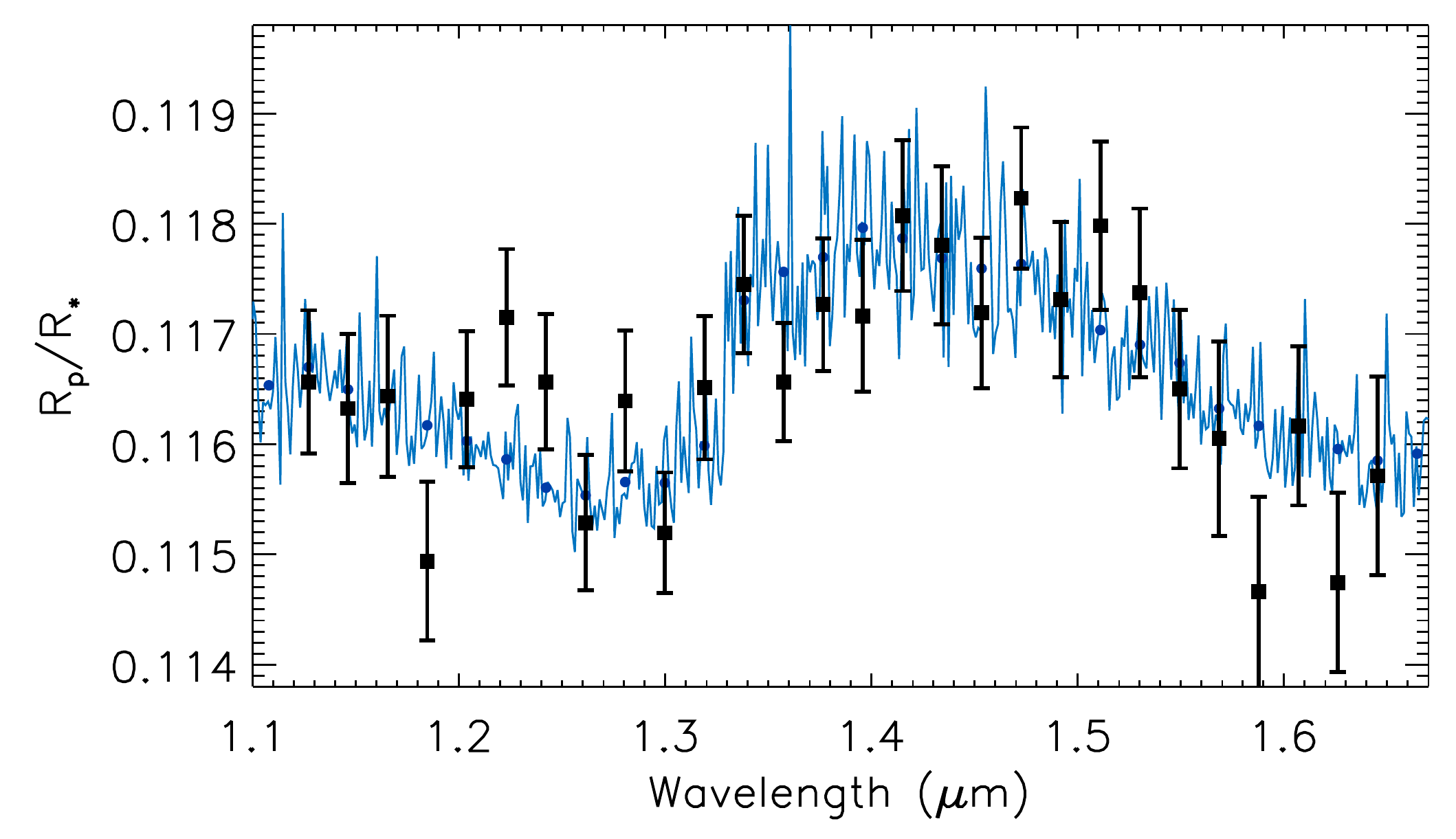}
	\end{center}
\caption{The transmission spectrum of HAT-P1b, using differential photometry with individual parameter fitting (see Fig 9a).  The full resolution planetary-averaged HAT-P-1b specific model is plotted in blue (based on the Fortney et al. 2008,2010 models).}
\label{fig:fullmodel}
\end{figure}

	The models based on Fortney et al. (2008; 2010) included a self-consistent treatment of radiative transfer and thermo-chemical equilibrium of neutral and ionic species.  The models assumed a solar metallically and local thermo-chemical equilibrium, accounting for condensation and thermal ionisation though no photochemistry (\citealt{lodders1999}; \citealt{loddersfegley2002}; \citealt{lodders2002}; \citealt{visscher2006}, \citealt{loddersfegley2006}; \citealt{lodders2009}; \citealt{freedman2008}).  
	In addition to isothermal models, transmission spectra were calculated using 1D temperature-pressure (T-P) profiles for the dayside, as well as an overall cooler planetary-averaged profile.  Models were also generated both with and without the inclusion of TiO and VO opacities.\\
	\indent{} The models based on \citet{burrows2010} and \citet{howe2012} used a 1D dayside T-P profile with stellar irradiation, in radiative, chemical, and hydrostatic equilibrium.  Chemical mixing ratios and corresponding opacities assume solar metallicity and local thermodynamical chemical equilibrium accounting for condensation with no ionisation, using the opacity database from \citet{sharpburrows2007} and the equilibrium chemical abundances from \citet{burrows1999} and \citet{burrows2001}. \\
	\indent{} \emph{Isothermal models:} Comparison of the observed atmospheric features to those produced by isothermal hydrostatic uniform abundance models helps provide an overall understanding of the observed features and any departures from them. We used isothermal models for T$_{eff}$=1500\,K (to represent the hotter dayside) for model atmospheres with and without TiO/VO and for a cooler isothermal model at T$_{eff}$=1000\,K (to represent the cooler terminator). The near-IR transit spectrum is relatively insensitive to the presence of TiO and VO. Models at T$_{eff}$=1500\,K including or not TiO/VO provided a poor fit with a $\chi^2$ value of  $\sim$54.5 for 27 degrees of freedom (DOF)  and can be rejected with a greater than 3$\sigma$ confidence. The T$_{eff}$=1000\,K model yielded an improved fit with a $\chi^{2}$ value of 35.68 for 27 DOF (see Fig. \ref{fig:model}).\\
	\indent{} \emph{HAT-P-1b specific models:} We also compared the transit spectrum to the transmission spectra generated by both a planetary averaged T-P profile and a dayside-averaged T-P profile specifically generated for HAT-P-1b. The model using the cooler planetary averaged T-P profile is our best fitting model giving a $\chi^2$ value of 26.89 for 27 DOF, while the hotter dayside-averaged T-P profile gives a marginally worse fit with a $\chi^2$ value of 28.87 for 27 DOF. We also compared the HAT-P-1b dayside model without TiO/VO from \citet{burrows2013}, and found a $\chi^2$ value of 37.68 for 27 DOF. 
	While this is a better fit than with the 1500\,K isothermal model, the cooler planetary averaged T-P profile and 1000\,K isothermal model have a stronger correlation to the data (see Fig. \ref{fig:model}). \\
	\indent{} To determine the overall significance of the model fits, we also calculated the fit for a straight line through the average planetary radius, corresponding to the case where no atmospheric features are detected. This gave a $\chi^2$ value of 56.71 for 27 DOF. Thus, we can rule out the null hypothesis at the 5.4 �sigma significance level, compared to our best-fitting atmospheric model using a planetary averaged T-P profile (see Fig.\ref{fig:fullmodel}).
	
\subsubsection{Single target model fitting}
\indent{} In addition to the above analysis of the transmission spectrum shown in Fig \ref{fig:alltranspec}a, we apply the $\chi^{2}$ test to compare the pre-calculated models to the transmission spectrum computed using single target photometry with individual parameter analysis (Fig. \ref{fig:alltranspec}b). Figure \ref{fig:model} shows the six models outlined in Sec. 4.1 fitted to the transmission spectrum for single target photometry, where the only fitting parameter is the base planetary radius, with $\Delta$R$_{p}$/R$_{*}$$\sim$\,0.001 lower for single target photometry. \\
\indent{} Similar to the fit in Sec. 4.1 the two T$_{eff}$ = 1500\,K models representing the hotter dayside show a poor fit to the data and can be rejected with greater than 97\% confidence. The remaining models, including the T$_{eff}$ = 1000\,K isothermal model representing the cooler terminator, show a greater significance of fit to the data with a significance of 4.4$\sigma$ over the null hypothesis. 
The model using the cooler planetary-averaged T-P profile is our best fitting model with a $\chi^{2}$ value of 27.10 for 27 DOF compared to a $\chi^{2}$ value of 46.5 for 27 DOF using a straight line through the average planetary radius representing a featureless atmosphere. \\ 
\indent{} To further corroborate these results against different analysis techniques, we determined the amplitude of the water feature in the data for each of the WFC3 transmission spectra shown in Fig. 9.  This was determined by scaling our best fitting atmospheric model to each of the four spectra.  The fitted scaling factor can change, particularly in analysis d) where it is lower, although the difference is not significant as there is much higher red noise in the other three analysis methods, making them less sensitive to the water absorption feature.  

\begin{figure}
\begin{center}
	\includegraphics[width=8cm]{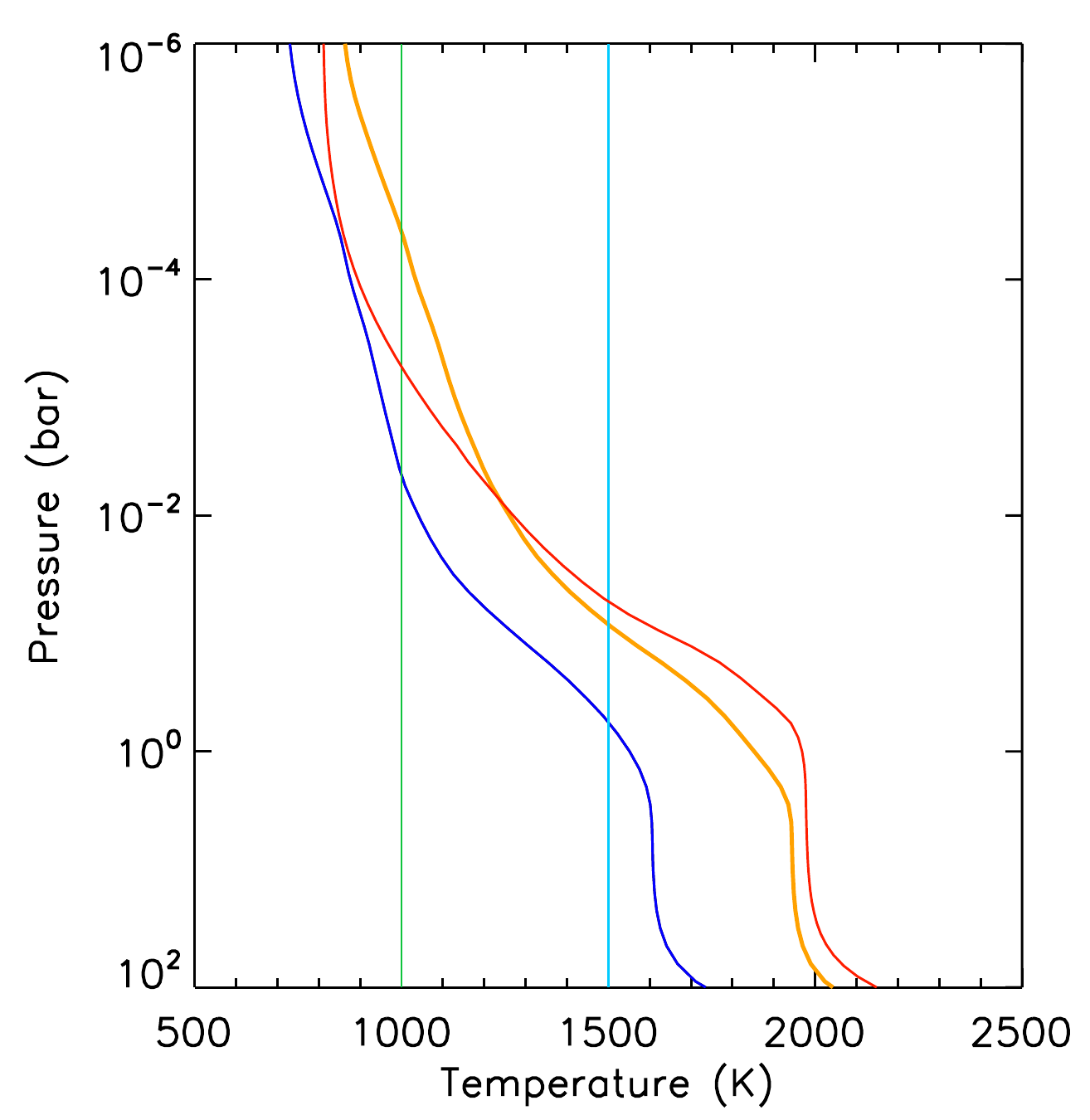}
	\end{center}
\caption{The temperature-pressure profile for the planetary-averaged profile (dark blue), the dayside-averaged profile (orange), and vertical lines marking the isothermal models at 1000\,K (green) and 1500\,K (light blue) (J. Fortney, 2012),  and the Burrows dayside model without TiO/VO (red).}
\label{fig:tp}
\end{figure}

\subsection{Implications for HAT-P-1b's structure:} 
	\indent{}  Given that transmission spectroscopy is mainly sensitive to the scale height, and therefore the absolute temperature of the atmosphere, we find evidence for a cooler temperature on average at the planetary limb, compared to the 1500\,K dayside brightness temperatures measured from Spitzer (\citealt{todorov2010}).  The 1000\,K isothermal model and the HAT-P-1b specific T-P profile models all show a significant improvement in the fit compared to a hotter 1500\,K isothermal model.  Therefore, a hotter temperature at lower pressures can be confidently ruled out.  This gives evidence that HAT-P-1b has cooler temperatures close to $\sim$1000\,K at $\sim$mbar pressures, where the best-fitting model T-P profiles overlap (see Fig. \ref{fig:tp}).  \\	
\indent{} The identification of atmospheric species is one of the first steps for understanding the nature of exoplanetary atmospheres. The presence of key species, or the lack there of, provides information on the exoplanets composition, chemistry, temperature, and atmospheric structures such as clouds or hazes; thus helping us place exoplanets into subcategories. Recent 3D hot-Jupiter models have shown that the warmer dayside temperatures can increase the atmospheric scale height and effectively ``puff-up" the dayside atmosphere, obscuring the cooler planetary limb as well as nightside spectral signatures (\citealt{fortney2010}). Although there is a difference of 1.5$\sigma$ between the warmer dayside-averaged T-P profile and that of the cooler planetary-averaged profile, the hotter model cannot be rejected with enough confidence to entirely rule it out and determine if the dayside atmosphere is significantly ``puffed-up" in the presence of high stellar irradiation.
The derived water feature is expected to be at a pressure of roughly 20 mbar at solar abundances (see Fig. \ref{fig:tp}). 
The derived water feature displays a similar amplitude to that seen in WASP-19b (\citealt{huitson2013}) with both planets consistent with a H$_{2}$O dominated atmospheric transmission in the near-IR. These observations show a contrast to HD\,209458b and XO-1b (\citealt{deming2013}), which both appear muted in water absorption, by perhaps cloud or haze, demonstrating a range in the presence of water in hot Jupiter atmospheres.

%
\section{Conclusion}
In this paper we present new measurements of HAT-P-1b's transmission spectrum using HST/WFC3 in spatial scan mode with precisions of $\sigma_{R_{p}/R_{*}} \simeq$ 0.00069 reached in 28 simultaneously measured wavelength bins. We find evidence for H$_2$O absorption in the atmosphere at 1.4\,$\mu$m with a greater than 5$\sigma$ significance level, with models in favour of a cooler planetary-averaged T-P profile at the limb of the planet near $\sim$millibar pressures for both single target and differential photometry. The amplitude of the derived water absorption is consistent with a H$_{2}$O dominated atmospheric transmission in the near-IR with evidence for a non-inverted T-P profile. The 1000\,K isothermal models show a significant improvement over hotter 1500\,K isothermal models, however, a ``puffed-up" dayside cannot be ruled out.  \\
\indent{} In our spatially scanned data, we find that performing differential photometry with individual parameter fitting of HST phase to the 7th order and removal of residual white-light common mode trends produces the best results, though the spectral shape is fairly independent of the different data reduction processes.  The use of spatial scan mode allowed us to take longer exposures therefore increasing the number of detected photons before saturation occurs, and reducing the effect of non-linearity and persistence in the IR detector. The spatial-scan mode allowed us to obtain the transmission spectrum of HAT-P-1b  at the resolution of the instrument at precessions equivalent to about one scale height of the planets' atmosphere per bin. As HAT-P-1 is also a member of a binary star system we were also able to use the resolved companion as a reference star to perform differential photometry,  removing some systematics and reducing the errors of the observations. This allowed for increasing the resolution of the measurements without significantly increasing the errors.  \\
\indent{} Future observations with our program using WFC3 in spatial scan mode will be able to better explore the diversity of H$_{2}$O in the atmospheres of close-in giant planets.

%
\section{Acknowledgements} 
H.R. Wakeford and D.K. Sing acknowledge support from STFC. All US-based co-authors acknowledge support from the Space Telescope Science Institute under HST-GO-12473 grants to their respective institutions. This work is based on observations with the NASA/ESA Hubble Space Telescope. This research has made use of NASA�s Astrophysics Data System, and components of the IDL astronomy library. 
We thank the referee for their useful comments. 

%
\footnotesize{
\bibliographystyle{mn2e}
\bibliography{paperlib}

\begin{thebibliography}{59}
\expandafter\ifx\csname natexlab\endcsname\relax\def\natexlab#1{#1}\fi

\bibitem[{Bakos {et~al}\mbox{.}(2011)Bakos, Hartman, Torres, Latham,
  Kov{\'a}cs, Noyes, Fischer, Johnson, Marcy, Howard, Kipping, Esquerdo,
  Shporer, B{\'e}ky, Buchhave, Perumpilly, Everett, Sasselov, Stefanik,
  L{\'a}z{\'a}r, Papp, \& S{\'a}ri}]{bakos2011}
Bakos G.~{\'A}. {et~al.}, 2011, \apj, 742, 116

\bibitem[{{Bakos} {et~al}\mbox{.}(2007){Bakos}, {Noyes}, {Kov{\'a}cs},
  {Latham}, {Sasselov}, {Torres}, {Fischer}, {Stefanik}, {Sato}, {Johnson},
  {P{\'a}l}, {Marcy}, {Butler}, {Esquerdo}, {Stanek}, \& et~al.}]{bakos2007}
{Bakos} G.~{\'A}. {et~al.}, 2007, \apj, 656, 552

\bibitem[{{Ben-Jaffel} \& {Ballester}(2013)}]{benjaffel2013}
{Ben-Jaffel} L., {Ballester} G.~E., 2013, \aap, 553, A52

\bibitem[{Berta {et~al}\mbox{.}(2012)Berta, Charbonneau, D{\'e}sert, Kempton,
  McCullough, Burke, Fortney, Irwin, Nutzman, \& Homeier}]{berta2012}
Berta Z.~K. {et~al.}, 2012, \apj, 747, 35

\bibitem[{{Brown}(2001)}]{brown2001a}
{Brown} T.~M., 2001, \apj, 553, 1006

\bibitem[{Brown {et~al}\mbox{.}(2001)Brown, Charbonneau, Gilliland, Noyes, \&
  Burrows}]{brown2001b}
Brown T.~M., Charbonneau D., Gilliland R.~L., Noyes R.~W., Burrows A., 2001,
  \apj, 552, 699

\bibitem[{Burrows {et~al}\mbox{.}(2001)Burrows, Hubbard, Lunine, \&
  Liebert}]{burrows2001}
Burrows A., Hubbard W.~B., Lunine J.~I., Liebert J., 2001, Rev. Mod. Phys., 73,
  719

\bibitem[{Burrows {et~al}\mbox{.}(2010)Burrows, Rauscher, Spiegel, \&
  Menou}]{burrows2010}
Burrows A., Rauscher E., Spiegel D.~S., Menou K., 2010, \apj, 719, 341

\bibitem[{{Burrows} \& {Sharp}(1999)}]{burrows1999}
{Burrows} A., {Sharp} C.~M., 1999, \apj, 512, 843

\bibitem[{{Burrows, A.}(2013)}]{burrows2013}
{Burrows, A.}, 2013, Atmospheric models for the hot jupiter hat-p-1b, to be
  published in the Exoplanets Special Feature of the PNAS

\bibitem[{{Butler} {et~al}\mbox{.}(2006){Butler}, {Wright}, {Marcy}, {Fischer},
  {Vogt}, {Tinney}, {Jones}, {Carter}, {Johnson}, {McCarthy}, \&
  {Penny}}]{butler2006}
{Butler} R.~P. {et~al.}, 2006, \apj, 646, 505

\bibitem[{Charbonneau {et~al}\mbox{.}(2002)Charbonneau, Brown, Noyes, \&
  Gilliland}]{charbonneau2002}
Charbonneau D., Brown T.~M., Noyes R.~W., Gilliland R.~L., 2002, \apj, 568, 377

\bibitem[{{de Mooij} {et~al}\mbox{.}(2011){de Mooij}, {de Kok}, {Nefs}, \&
  {Snellen}}]{demooij2011}
{de Mooij} E.~J.~W., {de Kok} R.~J., {Nefs} S.~V., {Snellen} I.~A.~G., 2011,
  \aap, 528, A49

\bibitem[{{Deming} {et~al}\mbox{.}(2013){Deming}, {Wilkins}, {McCullough},
  {Burrows}, {Fortney}, {Agol}, {Dobbs-Dixon}, {Madhusudhan}, {Crouzet},
  {Desert}, {Gilliland}, {Haynes}, {Knutson}, {Line}, {Magic}, \&
  {Mandell}}]{deming2013}
{Deming} D. {et~al.}, 2013, ArXiv:1302.1141

\bibitem[{D{\'e}sert {et~al}\mbox{.}(2009)D{\'e}sert, des Etangs, H{\'e}brard,
  Sing, Ehrenreich, Ferlet, \& Vidal-Madjar}]{desert2009}
D{\'e}sert J.-M., des Etangs A.~L., H{\'e}brard G., Sing D.~K., Ehrenreich D.,
  Ferlet R., Vidal-Madjar A., 2009, \apj, 699, 478

\bibitem[{Dressel {et~al}\mbox{.}(2010)Dressel, Wong, Pavlovsky, Long,
  {et~al.}}]{dressel2010}
Dressel L., Wong M., Pavlovsky C., Long K., {et~al.}, 2010, Wide field camera 3
  instrument handbook

\bibitem[{Eastman {et~al}\mbox{.}(2013)Eastman, Gaudi, \& Agol}]{eastman2012a}
Eastman J., Gaudi B.~S., Agol E., 2013, \pasp, 125, 83

\bibitem[{{Fortney}(2005)}]{fortney2005}
{Fortney} J.~J., 2005, \mnras, 364, 649

\bibitem[{Fortney {et~al}\mbox{.}(2010)Fortney, Shabram, Showman, Lian,
  Freedman, Marley, \& Lewis}]{fortney2010}
Fortney J.~J., Shabram M., Showman A.~P., Lian Y., Freedman R.~S., Marley
  M.~S., Lewis N.~K., 2010, \apj, 709, 1396

\bibitem[{{Freedman} {et~al}\mbox{.}(2008){Freedman}, {Marley}, \&
  {Lodders}}]{freedman2008}
{Freedman} R.~S., {Marley} M.~S., {Lodders} K., 2008, \apjs, 174, 504

\bibitem[{{Gibson} {et~al}\mbox{.}(2012){Gibson}, {Aigrain}, {Pont}, {Sing},
  {D{\'e}sert}, {Evans}, {Henry}, {Husnoo}, \& {Knutson}}]{gibson2012b}
{Gibson} N.~P. {et~al.}, 2012, \mnras, 422, 753

\bibitem[{{Grillmair} {et~al}\mbox{.}(2008){Grillmair}, {Burrows},
  {Charbonneau}, {Armus}, {Stauffer}, {Meadows}, {van Cleve}, {von Braun}, \&
  {Levine}}]{grillmair2008}
{Grillmair} C.~J. {et~al.}, 2008, \nat, 456, 767

\bibitem[{{Hayek} {et~al}\mbox{.}(2012){Hayek}, {Sing}, {Pont}, \&
  {Asplund}}]{hayek2012}
{Hayek} W., {Sing} D., {Pont} F., {Asplund} M., 2012, \aap, 539, A102

\bibitem[{Howe \& Burrows(2012)}]{howe2012}
Howe A.~R., Burrows A.~S., 2012, \apj, 756, 176

\bibitem[{{Huitson} {et~al}\mbox{.}(2013){Huitson}, {Sing}, {Pont}, {Fortney},
  {Burrows}, {Wilson}, {Ballester}, {Nikolov}, {Gibson}, {Deming}, {Aigrain},
  {Evans}, {Henry}, {Lecavelier des Etangs}, {Showman}, {Vidal-Madjar}, \&
  {Zahnle}}]{huitson2013}
{Huitson} C.~M. {et~al.}, 2013, ArXiv:1307.2083

\bibitem[{Johnson {et~al}\mbox{.}(2008)Johnson, Winn, Narita, Enya, Williams,
  Marcy, Sato, Ohta, Taruya, Suto, Turner, Bakos, Butler, Vogt, Aoki, Tamura,
  Yamada, Yoshii, \& Hidas}]{johnson2008}
Johnson J.~A. {et~al.}, 2008, \apj, 686, 649

\bibitem[{Kuntschner {et~al}\mbox{.}(2009)Kuntschner, Bushouse, K{\"u}mmel, \&
  Walsh}]{kuntschner2009}
Kuntschner H., Bushouse H., K{\"u}mmel M., Walsh J., 2009, Wfc3 smov proposal
  11552: Calibration of the g141 grism. Tech. rep., MAST

\bibitem[{{Lecavelier des Etangs} {et~al}\mbox{.}(2012){Lecavelier des Etangs},
  {Bourrier}, {Wheatley}, {Dupuy}, {Ehrenreich}, {Vidal-Madjar}, {H{\'e}brard},
  {Ballester}, {D{\'e}sert}, {Ferlet}, \& {Sing}}]{lecavelier2012}
{Lecavelier des Etangs} A. {et~al.}, 2012, \aap, 543, L4

\bibitem[{Linsky {et~al}\mbox{.}(2010)Linsky, Yang, France, Froning, Green,
  Stocke, \& Osterman}]{linsky2010}
Linsky J.~L., Yang H., France K., Froning C.~S., Green J.~C., Stocke J.~T.,
  Osterman S.~N., 2010, \apj, 717, 1291

\bibitem[{{Lodders}(1999)}]{lodders1999}
{Lodders} K., 1999, \apj, 519, 793

\bibitem[{{Lodders}(2002)}]{lodders2002}
{Lodders} K., 2002, \apj, 577, 974

\bibitem[{{Lodders}(2009)}]{lodders2009}
{Lodders} K., 2009, ArXiv: 0910.0811

\bibitem[{{Lodders} \& {Fegley}(2002)}]{loddersfegley2002}
{Lodders} K., {Fegley} B., 2002, \icarus, 155, 393

\bibitem[{{Lodders} \& {Fegley}(2006)}]{loddersfegley2006}
{Lodders} K., {Fegley}, Jr. B. edited by~{Mason} J.~W., 2006, {Chemistry of Low
  Mass Substellar Objects}, Springer Praxis, p.~1

\bibitem[{Madhusudhan(2012)}]{madhusudhan2012}
Madhusudhan N., 2012, \apj, 758, 36

\bibitem[{{Mandel} \& {Agol}(2002)}]{mandelagol2002}
{Mandel} K., {Agol} E., 2002, \apjl, 580, L171

\bibitem[{{Markwardt}(2009)}]{markwardt2009}
{Markwardt} C.~B., 2009, in Astronomical Society of the Pacific Conference
  Series, Vol. 411, Astronomical Data Analysis Software and Systems XVIII,
  {Bohlender} D.~A., {Durand} D., {Dowler} P., eds., p. 251

\bibitem[{{McCullough}(2011)}]{mccullough2011}
{McCullough} P., 2011, Wfc space telescope analysis newsletter 6

\bibitem[{Moses {et~al}\mbox{.}(2013)Moses, Madhusudhan, Visscher, \&
  Freedman}]{moses2012}
Moses J., Madhusudhan N., Visscher C., Freedman R., 2013, \apj, 763, 25

\bibitem[{{Narita} {et~al}\mbox{.}(2005){Narita}, {Suto}, {Winn}, {Turner},
  {Aoki}, {Leigh}, {Sato}, {Tamura}, \& {Yamada}}]{narita2005}
{Narita} N. {et~al.}, 2005, \pasj, 57, 471

\bibitem[{{Nikolov} {et~al}\mbox{.}(2013){Nikolov}, {Sing}, {Pont}, {Burrows},
  {Fortney}, {Ballester}, {Evans}, {Huitson}, {Wakeford}, {Wilson}, {Aigrain},
  {Deming}, \& {Gibson}}]{nikolov2013}
{Nikolov} N. {et~al.}, 2013, \mnras,~ submitted

\bibitem[{{Pont} {et~al}\mbox{.}(2007){Pont}, {Gilliland}, {Moutou},
  {Charbonneau}, {Bouchy}, {Brown}, {Mayor}, {Queloz}, {Santos}, \&
  {Udry}}]{pont2007}
{Pont} F. {et~al.}, 2007, \aap, 476, 1347

\bibitem[{{Pont} {et~al}\mbox{.}(2013){Pont}, {Sing}, {Gibson}, {Aigrain},
  {Henry}, \& {Husnoo}}]{pont2012}
{Pont} F., {Sing} D.~K., {Gibson} N.~P., {Aigrain} S., {Henry} G., {Husnoo} N.,
  2013, \mnras, 432, 2917

\bibitem[{{Pont} {et~al}\mbox{.}(2006){Pont}, {Zucker}, \& {Queloz}}]{pont2006}
{Pont} F., {Zucker} S., {Queloz} D., 2006, \mnras, 373, 231

\bibitem[{{Redfield} {et~al}\mbox{.}(2008){Redfield}, {Endl}, {Cochran}, \&
  {Koesterke}}]{redfield2008}
{Redfield} S., {Endl} M., {Cochran} W.~D., {Koesterke} L., 2008, \apjl, 673,
  L87

\bibitem[{{Seager} \& {Sasselov}(2000)}]{seager2000}
{Seager} S., {Sasselov} D.~D., 2000, \apj, 537, 916

\bibitem[{{Sharp} \& {Burrows}(2007)}]{sharpburrows2007}
{Sharp} C.~M., {Burrows} A., 2007, \apjs, 168, 140

\bibitem[{{Sing}(2010)}]{sing2010}
{Sing} D.~K., 2010, \aap, 510, A21

\bibitem[{{Sing} {et~al}\mbox{.}(2011){Sing}, {Pont}, {Aigrain}, {Charbonneau},
  {D{\'e}sert}, {Gibson}, {Gilliland}, {Hayek}, {Henry}, {Knutson}, {Lecavelier
  Des Etangs}, {Mazeh}, \& {Shporer}}]{sing2011b}
{Sing} D.~K. {et~al.}, 2011, \mnras, 416, 1443

\bibitem[{{Snellen} {et~al}\mbox{.}(2008){Snellen}, {Albrecht}, {de Mooij}, \&
  {Le Poole}}]{snellen2008}
{Snellen} I.~A.~G., {Albrecht} S., {de Mooij} E.~J.~W., {Le Poole} R.~S., 2008,
  \aap, 487, 357

\bibitem[{{Swain} {et~al}\mbox{.}(2013){Swain}, {Deroo}, {Tinetti}, {Hollis},
  {Tessenyi}, {Line}, {Kawahara}, {Fujii}, {Showman}, \&
  {Yurchenko}}]{swain2012}
{Swain} M. {et~al.}, 2013, \icarus, 225, 432

\bibitem[{Swain {et~al}\mbox{.}(2008)Swain, Vasisht, \& Tinetti}]{swain2008}
Swain M.~R., Vasisht G., Tinetti G., 2008, Nature, 452, 329

\bibitem[{{Tinetti} {et~al}\mbox{.}(2007){Tinetti}, {Vidal-Madjar}, {Liang},
  {Beaulieu}, {Yung}, {Carey}, {Barber}, {Tennyson}, {Ribas}, {Allard},
  {Ballester}, {Sing}, \& {Selsis}}]{tinetti2007}
{Tinetti} G. {et~al.}, 2007, \nat, 448, 169

\bibitem[{{Todorov} {et~al}\mbox{.}(2010){Todorov}, {Deming}, {Harrington},
  {Stevenson}, {Bowman}, {Nymeyer}, {Fortney}, \& {Bakos}}]{todorov2010}
{Todorov} K., {Deming} D., {Harrington} J., {Stevenson} K.~B., {Bowman} W.~C.,
  {Nymeyer} S., {Fortney} J.~J., {Bakos} G.~A., 2010, \apj, 708, 498

\bibitem[{{Torres} {et~al}\mbox{.}(2008){Torres}, {Winn}, \&
  {Holman}}]{torres2008}
{Torres} G., {Winn} J.~N., {Holman} M.~J., 2008, \apj, 677, 1324

\bibitem[{Vidal-Madjar {et~al}\mbox{.}(2004)Vidal-Madjar, D{\'e}sert, des
  Etangs, H{\'e}brard, Ballester, Ehrenreich, Ferlet, McConnell, Mayor, \&
  Parkinson}]{vidal2004}
Vidal-Madjar A. {et~al.}, 2004, \apjl, 604, L69

\bibitem[{{Vidal-Madjar} {et~al}\mbox{.}(2003){Vidal-Madjar}, {Lecavelier des
  Etangs}, {D{\'e}sert}, {Ballester}, {Ferlet}, {H{\'e}brard}, \&
  {Mayor}}]{vidal2003}
{Vidal-Madjar} A., {Lecavelier des Etangs} A., {D{\'e}sert} J.-M., {Ballester}
  G.~E., {Ferlet} R., {H{\'e}brard} G., {Mayor} M., 2003, \nat, 422, 143

\bibitem[{{Visscher} {et~al}\mbox{.}(2006){Visscher}, {Lodders}, \&
  {Fegley}}]{visscher2006}
{Visscher} C., {Lodders} K., {Fegley}, Jr. B., 2006, \apj, 648, 1181

\bibitem[{Waldmann {et~al}\mbox{.}(2013)Waldmann, Tinetti, Deroo, Hollis,
  Yurchenko, \& Tennyson}]{waldmann2013}
Waldmann I.~P., Tinetti G., Deroo P., Hollis M.~D., Yurchenko S.~N., Tennyson
  J., 2013, \apj, 766, 7

\end{thebibliography}
}

\label{lastpage}
\end{document}